\documentclass[12pt]{article}
\usepackage{eurosym}
\usepackage[left=0.5in, right=0.5in, top=0.75in, bottom=0.75in]{geometry}
\usepackage{hyperref}
\usepackage{mathtools}
\usepackage{graphicx}
\usepackage{float}
\usepackage{blindtext}
\usepackage{mathrsfs}
\usepackage{pstricks}
\usepackage{amsmath}
\usepackage{multirow}
\usepackage{booktabs,tabularx}
\newcolumntype{C}{>{\centering\arraybackslash}X}
\usepackage[figurename=Fig.]{caption}

\begin{document}

	\begin{center}
	\textit{\large Wigner function of noisy accelerated two-qubit system}\\
	\vspace{0.5cm}
	M. Y. Abd-Rabbou$^{a}$\textit{\footnote{e-mail:m.elmalky@azhar.edu.eg}}
	,N. Metwally$^{b,c}$\textit{\footnote{Nmetwally@aswu.edu.eg}} , M. M. A. Ahmed $^{a}$, and  A.-S. F. Obada $^{a}$,
	
	$^{a}${\footnotesize Mathematics Department, Faculty of Science, Al-Azhar
		University, Nasr City 11884, Cairo.}
	
	$^{b}${\footnotesize Math. Dept., College of Science, University of Bahrain, Bahrain.}
	
	$^{c}${\footnotesize Department of Mathematics, Aswan University
		Aswan, Sahari 81528, Egypt}
\end{center}

\begin{abstract}
In this manuscript, the behavior of the Wigner function of accelerated and non-accelerated two qubit system  passing through different noisy channels is discussed. The decoherence of the initial quantum correlation due to the noisy channels and the acceleration process is investigated by means of Wigner function.  The negative (positive) behavior of the Wigner function predicts  the gain of the quantum (classical) correlations. Based on the upper and lower bounds of the Wigner function, the entangled initial state loses its quantum correlation due the acceleration process and the strengths of the noisy channels.  However, by controlling the distribution angles, the decoherence of these quantum correlation may be suppressed.  For accelerated state, the robustness of the quantum correlations contained in the initial state appears in different ranges of the distribution angles depending on the noisy type. For the bit phase flip and the phase flip channels, the  robustness of the quantum correlations is shown at any acceleration and large range of distribution angles. However, the fragility of the quantum correlation is depicted  for large  values for strength of the bit flip channel. Different profiles of the Wigner function are exhibited for the quantum and classical correlations, cup, lune, hemisphere.

\end{abstract}

\section{Introduction.}
 It is well known that the reconstruction of  density operator may be done  by the  quasi-probability distribution $(Q-PD)$ of the radiation field \cite{PhysRevA.48.2479}.  These types of distributions are used as predictors of   the  non-classicality  behavior of the  quantum state\cite{deleglise2008reconstruction,mcconnell2015entanglement,mohamed2018nonclassical}, where their  negative values are indicators of the existence  of the  quantum correlation. Due to their importance, there are many studies devoted to study them on different systems. For example, Ref. \cite{obada2012wigner} investigated analytically the Wigner function distribution of a two-qubit field system  in the presence of pure phase noisy. The $s$-parameterized $Q-PD$ is described in the  angular momentum basis via atomic coherent state \cite{biedenharn1984angular,klimov2017generalized}. However, the value of $s$-parameters determines the type of the $Q-PD$, where $s=-1,0,1$, represent the Husimi-Berezin $Q$-function, \cite{husimi1940some,PhysRevA.57.671},  Wigner quasi-distribution function\cite{varilly1989moyal,klimov20022}, and the $P$-function \cite{PhysRevLett.10.277}, respectively.

 However,  Wigner function has been used widely to  study  the phase space in continuous or discrete variable \cite{PhysRevA.70.062101}. The time evolution of superconducting flux qubits coupled to a system of electrons is analyzed by SU(2) Wigner function\cite{reboiro2015use}. The three qubit states has been reconstructed experimentally through  the Wigner distribution function  \cite{ciampini2017wigner}. A framework for representing any general quantum state of arbitrary finite-dimension as a complete continuous Wigner function has been presented \cite{PhysRevLett.117.180401,koczor2018time}.

  As far as we know, the Wigner function of  accelerated  quantum systems  was  not discussed widely. Therefore, we are motivated to study it for accelerated two qubit systems. Moreover,  the effect of different noisy channels on the behavior of the Wigner function is discussed. We employ the  behavior of the Wigner function as a predictor of the classical and the quantum correlations, where we investigate the effect of the noisy channels strengths as well as the acceleration parameter on the quantum and classical correlations.

The layout of this manuscript is as follows: in Sec.(\ref{s.3.2}), we introduce analytical forms of the quasi-probability distributions  Wigner function. The suggested model is introduced in Sec.(\ref{s.3.3}), where an analytical form of the Wigner function of the non-accelerated system is obtained. Sec.(\ref{s.3.4}), is devoted   to discuss the behavior of  the Wigner function for accelerated system. The effect of the amplitude, bit-phase flip, bit flip and phase noisy channels on the behavior of the Wigner function is discussed in  Sec.(\ref{s.3.5}). Finally, we summarize our results in Sec.(\ref{s.3.6}).

\section{Formalism of $  SU(2) $ Quasi-Distribution.}\label{s.3.2}

 The $ r $-parameterized family of quasi-probability distributions (Q-PD) in SU(2) algebra are reconstructed by the standard angular momentum basis $ |m,S\rangle  , m=-S,...,S $ as follows\cite{PhysRevA.24.2889,metwally2019wigner}:
 \begin{equation}\label{3.1}
 	W^{(r)}_{\hat{\rho}}(\theta,\phi)=Tr[\hat{\rho}_{a,b}\hat{A}_a^{(r)}(\theta,\phi)\hat{A}_b^{(r)}(\theta,\phi)],
 \end{equation}
where  $r=-1,0,1$ for the $ Q_{\hat{\rho}}(\theta,\phi) $, Wigner $W_{\hat{\rho}}(\theta,\phi) $  and the $ P_{\hat{\rho}}(\theta,\phi) $ functions, respectively. The operator $ \hat{A}^{(r)}(\theta,\phi) $ is defined by:
\begin{equation}\label{3.2}
	\hat{A}_i^{(r)}(\theta,\phi)=\sqrt{\frac{4 \pi}{2S+1}} \sum_{L_i=0}^{2S} \sum_{M=-L}^{L} (C^{S,S}_{S,S;L,0})^{-r} \hat{T} ^{(S_i)^{\dagger}}_{L,M} Y^i _{L,M}(\theta,\phi),
\end{equation}
where $ i $ refers to qubit $ a(b) $, and $Y^i_{L,M}(\theta,\phi) $ are the spherical harmonics functions, while $ T^{(S_i)^{\dagger}}_{L,M}= (-1)^{M}  T^{(S)}_{L,-M}$ are the orthogonal irreducible tensor operators which  are represented in $ (2S+1) $-dimensions Hilbert space as a linear combination by \cite{klimov20022}:

\begin{equation}\label{3.3}
 \hat{T}^{(S_i)^{\dagger}}_{L,M}=(-1)^{M}\sqrt{\frac{2L+1}{2S+1}} \sum_{m,m'=-S_i}^{S_i} C^{S_i, m'}_{S_i,m;L,-M}|S_i,m'\rangle\langle S_i,m |,
\end{equation}
the coefficient $ C^{S,m'}_{S,m;L,-M} $ is the Clebsch-Gordan coupling coefficient, where $ 0 \leq L_i \leq 2S $, and $ -L\leq M \leq L $.
The $ r $-parameterized Q-PD  at $ S=\frac{1}{2} $ is given by,
\begin{equation}\label{3.4}
\begin{split}
	W^{(r)}_{\hat{\rho}}(\theta,\phi)=2\pi \ Tr&\bigg[\hat{\rho}_{a,b}\big( \hat{T}^{(a)^{\dagger}}_{0,0} Y^a_{0,0}(\theta,\phi)+ (\sqrt{3})^{(r)} \sum_{n=-1}^{1} \hat{T}^{(a)^{\dagger}}_{0,n} Y^a_{0,n}(\theta,\phi)\big)\\& \times \big( \hat{T}^{(b)^{\dagger}}_{0,0} Y^b_{0,0}(\theta,\phi)+ (\sqrt{3})^{(r)} \sum_{n=-1}^{1} \hat{T}^{(b)^{\dagger}}_{0,n} Y^b_{0,n}(\theta,\phi)\big)\bigg],
\end{split}
\end{equation}
where,
\begin{equation*}
	\begin{split}
		&\hat{T}^{(i)^{\dagger}}_{0,0}=\frac{1}{\sqrt{2}}(|0\rangle_i \langle 0|+ |1\rangle_i \langle 1|), \quad \hat{T}^{(i)^{\dagger}}_{1,0}=\frac{-1}{\sqrt{2}}(|0\rangle_i \langle 0|- |1\rangle_i \langle 1|), \quad \hat{T}^{(i)^{\dagger}}_{1,-1}=|1\rangle_i \langle 0|\quad\\&
		\hat{T}^{(i)^{\dagger}}_{1,1}=-|0\rangle_i \langle 1|,\quad |1\rangle= |\frac{1}{2}, \frac{1}{2}\rangle=|\frac{-1}{2}, \frac{-1}{2}\rangle, \quad |0\rangle= |\frac{-1}{2}, \frac{1}{2}\rangle= |\frac{1}{2}, \frac{-1}{2}\rangle.
	\end{split}
\end{equation*}

\section{The Suggested Model.}\label{s.3.3}
 In this contribution, we assume that a system  of two qubits is initially prepared in the $X-$ state.  In the  set of the computational basis
 $\{ |00\rangle,  |01\rangle, |10\rangle, |11\rangle\}$, the density operator of the system is given by,

\begin{equation}\label{3.5}
\begin{split}
	\hat{\rho}_{ab}(0)=&(\varrho_{11}|0\rangle_a \langle 0|+\varrho_{22}|1\rangle_a \langle 1|) |0\rangle_b \langle 0|+ (\varrho_{33}|0\rangle_a \langle 0|+\varrho_{44}|1\rangle_a \langle 1|)|1\rangle_b \langle 1|\\&
	+(\varrho_{14}|0\rangle_a \langle 1|+\varrho_{23}|1\rangle_a \langle 0|) |0\rangle_b \langle 1|+ (\varrho_{32}|0\rangle_a \langle 1|+\varrho_{41}|1\rangle_a \langle 0|)|1\rangle_b \langle 0|,
\end{split}
\end{equation}
where
\begin{equation*}
	\begin{split}
	&\varrho_{11}=\varrho_{44}=\frac{1}{4}(1+c_3), \ \varrho_{22}=\varrho_{33}=\frac{1}{4}(1- c_3), \
	\varrho_{14}=\varrho_{41}=\frac{1}{4}(c_1-c_2),\ \varrho_{23}=\varrho_{32}=\frac{1}{4}(c_1+c_2)
	\end{split}
\end{equation*}
and $ c_i=Tr(\hat{\rho}_{ab}\sigma^a_i \sigma^b_i)$, $ \sigma^k_i $, $ i=1,2 \ \text{and}\ 3 $ are the Pauli spin matrices, while $ k$ indicates to  Alice's qubit $ a $ and Bob's qubit $ b $.
The main task of this contribution is investigating the behavior of the Wigner function when  only one qubit  accelerated. However, for the non-accelerated system, the Wigner function of the system (\ref{3.5}) is given by,
\begin{equation}\label{w0}
\begin{split}
	W_{\hat{\rho}}(\theta,\phi)&=2\pi \big[\varrho_{11} (\Psi^2_{11}+\Psi^2_{22})+\varrho_{14} (\Psi^2_{12}+\Psi^2_{21}) +2 \varrho_{22}\Psi_{11}\Psi_{22} +2\varrho_{23}\Psi_{12}\Psi_{21}\big],
\end{split}
\end{equation}
where the functions $\Psi_{ij}$ are defined by the following spherical harmonics,
\begin{equation*}
\begin{split}
&\Psi_{11}=\frac{1}{\sqrt{2}}(Y_{0,0}(\theta,\phi)-Y_{1,0}(\theta,\phi)),\ \
\Psi_{12}=-Y_{1,1}(\theta,\phi),\ \
\psi_{21}=Y_{1,-1}(\theta,\phi),\ \
\Psi_{22}=\frac{1}{\sqrt{2}}(Y_{0,0}(\theta,\phi)+Y_{1,0}(\theta,\phi)).
\end{split}
\end{equation*}

\section{Accelerated Winger Function.}\label{s.3.4}
\quad Now, let us assume that, Alice's qubit  is traveling with a uniform acceleration and  Bob's qubit remains in the inertial frame \cite{PhysRevA.82.042332}. In the  computational basis, $ |0_k\rangle$ and $  |1_k\rangle$ the Minkowski-Fock states are  transformed into  the Rindler-Fock states as \cite{PhysRevA.83.052306,metwally2017estimation}:
\begin{equation}\label{3.6}
	\begin{split}
	|0_k\rangle=\cos r |0_k\rangle_{I} |0_k\rangle_{II} + \sin r |1_k\rangle_{I} |1_k\rangle_{II} , \quad |1_k\rangle=a_k ^{\dagger} |0_k\rangle= |1_k\rangle_{I} |0_k\rangle_{II},
	\end{split}
\end{equation}
where $r$ is the acceleration such  that, $ \tan r = \exp(-\pi \omega c / a) $, $ 0\leq r \leq \pi/4 $, $ -\infty \leq a \leq \infty$, $ c $ is the speed of light, and $ \omega $ is the frequency.
Due to the transformation (\ref{3.6}), the space is splitting   into two regions, $I$  and $II$. By  tracing out over all the degrees of freedom on  the second region $II$, the final state which describes the accelerated system, $ \hat{\rho}^{acc}_{a,b} $ is given by,

\begin{equation}\label{3.7}
	\begin{split}
	\hat{\rho}^{acc}_{ab}=& \mathcal{A}_{11} |00\rangle \langle 00| + \mathcal{A}_{22} (|01\rangle \langle 01|+|10\rangle \langle 10|)+ \mathcal{A}_{33} |11\rangle \langle 11| + \mathcal{A}_{14} |00\rangle \langle 11|+(\mathcal{A}_{23} |10\rangle \langle 01|+h.c.).
	\end{split}
\end{equation}
 where
\begin{equation}
\begin{split}
&\mathcal{A}_{11}=\cos^4 r \varrho_{11},\ \mathcal{A}_{22}=\cos^2 r ( \sin^2r \varrho_{11}+\varrho_{22}),\ \ \mathcal{A}_{33}=\Bigl[(\sin^4 r+1)\varrho_{11}+ 2\sin^2r \varrho_{22}\Bigr],\ \mathcal{A}_{ij}=\cos ^2 r \varrho_{ij}, i \neq j.
\end{split}
\end{equation}

\begin{figure}[H]
	\centering
	\includegraphics[width=0.45\linewidth, height=5cm]{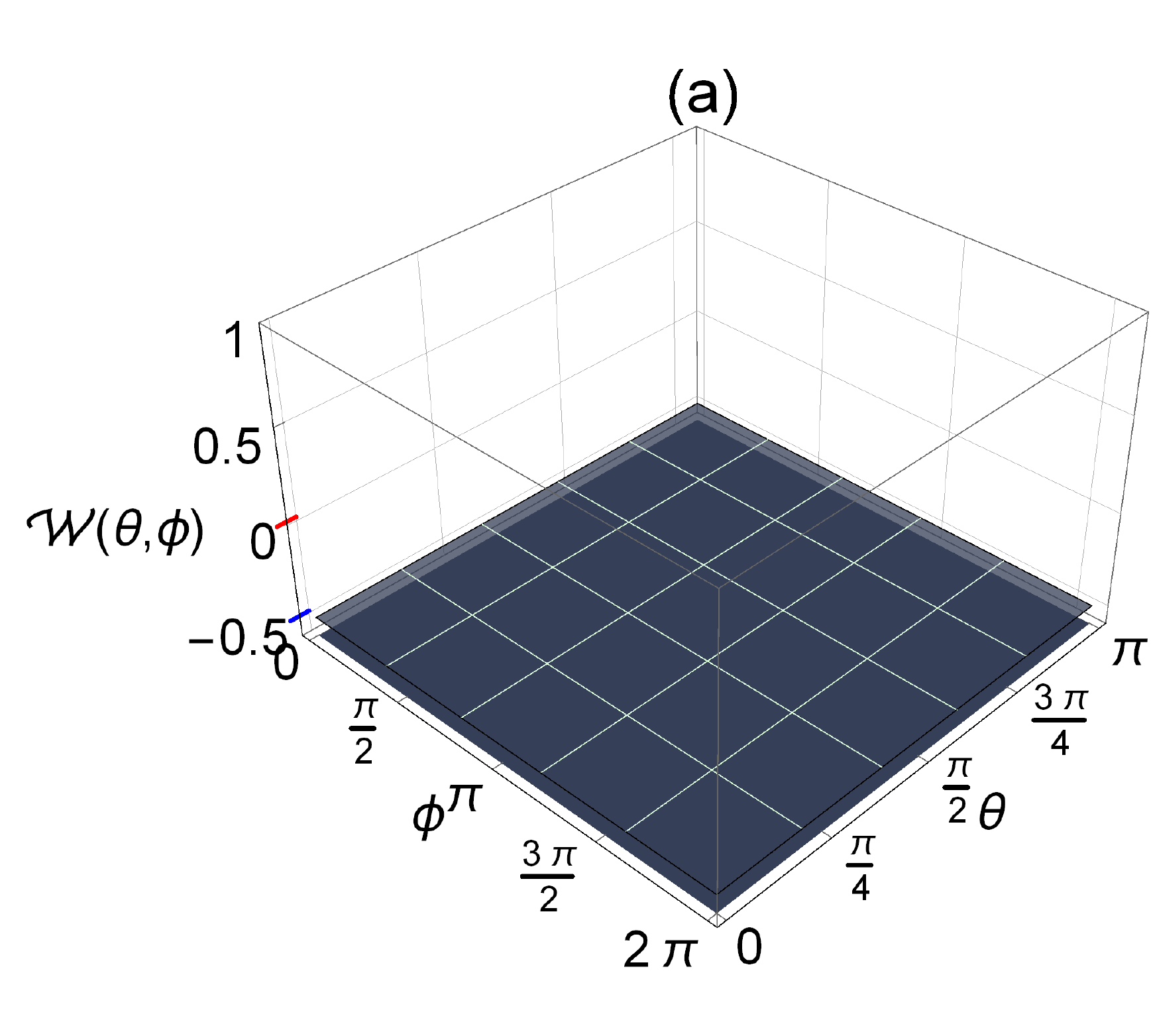}
	\includegraphics[width=0.45\linewidth, height=5cm]{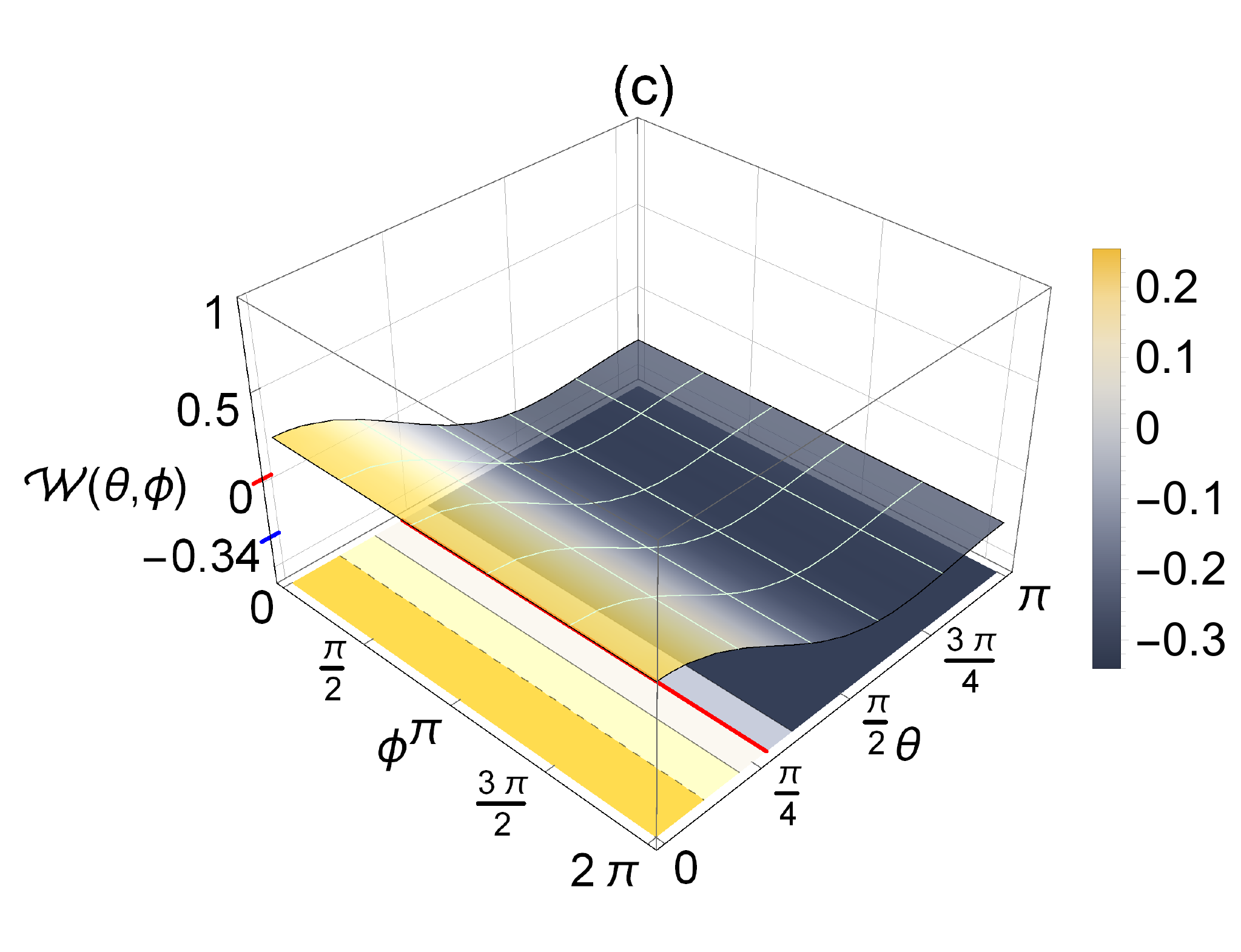}
	\caption{ The behavior of the Wigner function  $W(\theta~,\phi)$ of  a system is initially prepared in the singlet state, where (a)for the non-accelerated system  and (b) for the accelerated state with $ r=0.6$. }
	\label{fig:3.1}
\end{figure}

\begin{figure}[h!]
	\centering
	\includegraphics[width=0.9\linewidth, height=5cm]{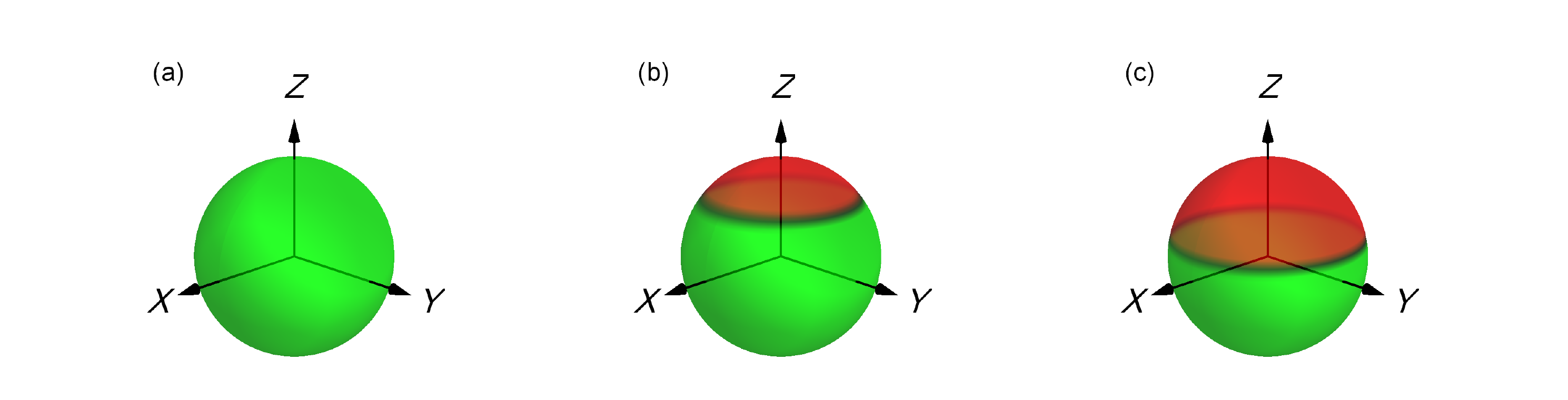}
	\caption{The behavior of $ W(\theta,\phi)$ on a sphere where  (a) $r=0 $, (b)   $ r=0.6$, (d)  $ r=0.78 $. }
	\label{fig:3.1.1}
\end{figure}

By using  Eqs.(\ref{3.4}) into Eq.(\ref{3.7}), one gets the  Wigner function of the two qubit system as:
\begin{equation}\label{wa}
\begin{split}
W_{\hat{\rho}^{acc}}(\theta,\phi)&=2\pi \big[\mathcal{A}_{11} \Psi^2_{11}+\mathcal{A}_{33}\Psi^2_{22}+2\mathcal{A}_{22}\Psi_{11}\Psi_{22} +\mathcal{A}_{14} (\Psi^2_{12}+\Psi^2_{21})+2\mathcal{A}_{23}\Psi_{12}\Psi_{21}\big].
\end{split}
\end{equation}

Figs.(\ref{fig:3.1}.a) and (\ref{fig:3.1}.b), display the behavior of the Wigner function, $ W(\theta,\phi)$ for a system initially prepared in the singlet state $\rho_{\psi^-}=|\psi^-\rangle\langle\psi^-|$. It is clear that, $W(\theta,\phi)<0$ for all values of the distribution angle $\theta$ and $\phi$. This predicts that, the system is completely entangled.
The behavior of the Wigner function(\ref{wa}), when only Alic's qubit is accelerated uniformly with $r=0.6$ is displayed in Fig.(\ref{fig:3.1}.b), where   the Wigner function decreases gradually as $\theta$ increases.  For the accelerated singlet state, the negative behavior of $ W_{\hat{\rho}^{acc}}$ is displayed at  small values of the distribution angle $\theta$. The phase parameter has  non-noticeable effect on the behavior of the Wigner function and consequently, the freezing phenomenon of the Wigner function is displayed, where the freezing degree depends on the parameter $\theta$.

Fig.(\ref{fig:3.1.1}), displays the Wigner functions behavior on a surface of a sphere at different values of the acceleration $r$, where the green area  indicates the negative behavior of the Wigner function, which means  the existence of the  quantum correlation. These results are  consistence with those  displayed in Fig.(\ref{fig:3.1}.a) at  $r=0$, where the green area is depicted in  $-1\leq z\leq 1$. However, as the acceleration increases, one can notice that a red  area appears, as upper cup with $\frac{1}{2}\leq z\leq 1$ which indicates that the  accelerated system loses its  quantum correlation and the classical correlations appear. However, the read area of the sphere increase by increasing the acceleration parameter $r=0.78$, where it is predicted in the region  $0\leq z\leq 1$.

\section{Noisy Channels Effect.} \label{s.3.5}
\quad Now, let us  assume that,  the accelerated Alice's qubit is forced to pass through  one of the noisy channels, which may be amplitude, phase,  bit-flip or phase-bit  channel \cite{PhysRevA.78.022322}. Mathematically, a suitable description of these channels is through the Kraus operators. However, the final output state may be given by \cite{PhysRevA.87.042108}:

\begin{equation}\label{ch.3}
\hat{\rho}^{ch}_{ab}=\sum_{i=}(E^a_i\otimes I_{2\times 2} )\hat{\rho}_{ab}(0)(E^{a^{\dagger}}_i\otimes I_{2\times 2} ),
\end{equation}
 where  $E^a_i $  are the Kraus operators  of  the used channel.

 \subsection{Amplitude damping channel $(\mathcal{C}_{ad})$.}
  For the amplitude damning channel ($\mathcal{C}_{ad}$), the Kraus operators $ E^a_i $   may be  defined as \cite{nielsen2002quantum}:
 \begin{equation}
 E_1=diag(1,\sqrt{1-p_{ad}}),\qquad E_2=\sqrt{p_{ad}}|0\rangle \langle 1|,
 \end{equation}
 where $ p_{ad} $ is the channel strength.
 \begin{figure}[H]
 	\centering
 	\includegraphics[width=0.45\linewidth, height=5cm]{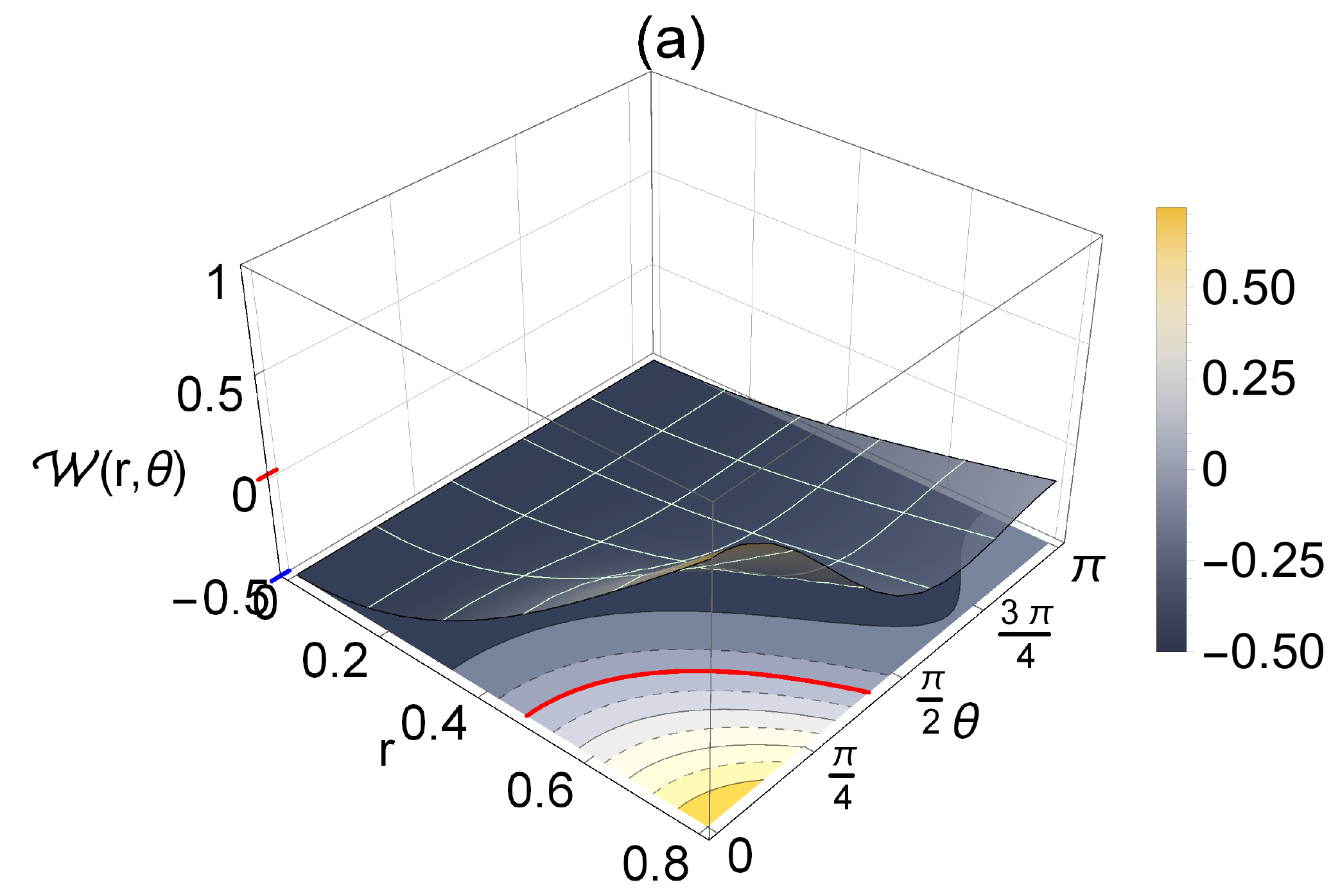}
 	\includegraphics[width=0.45\linewidth, height=5cm]{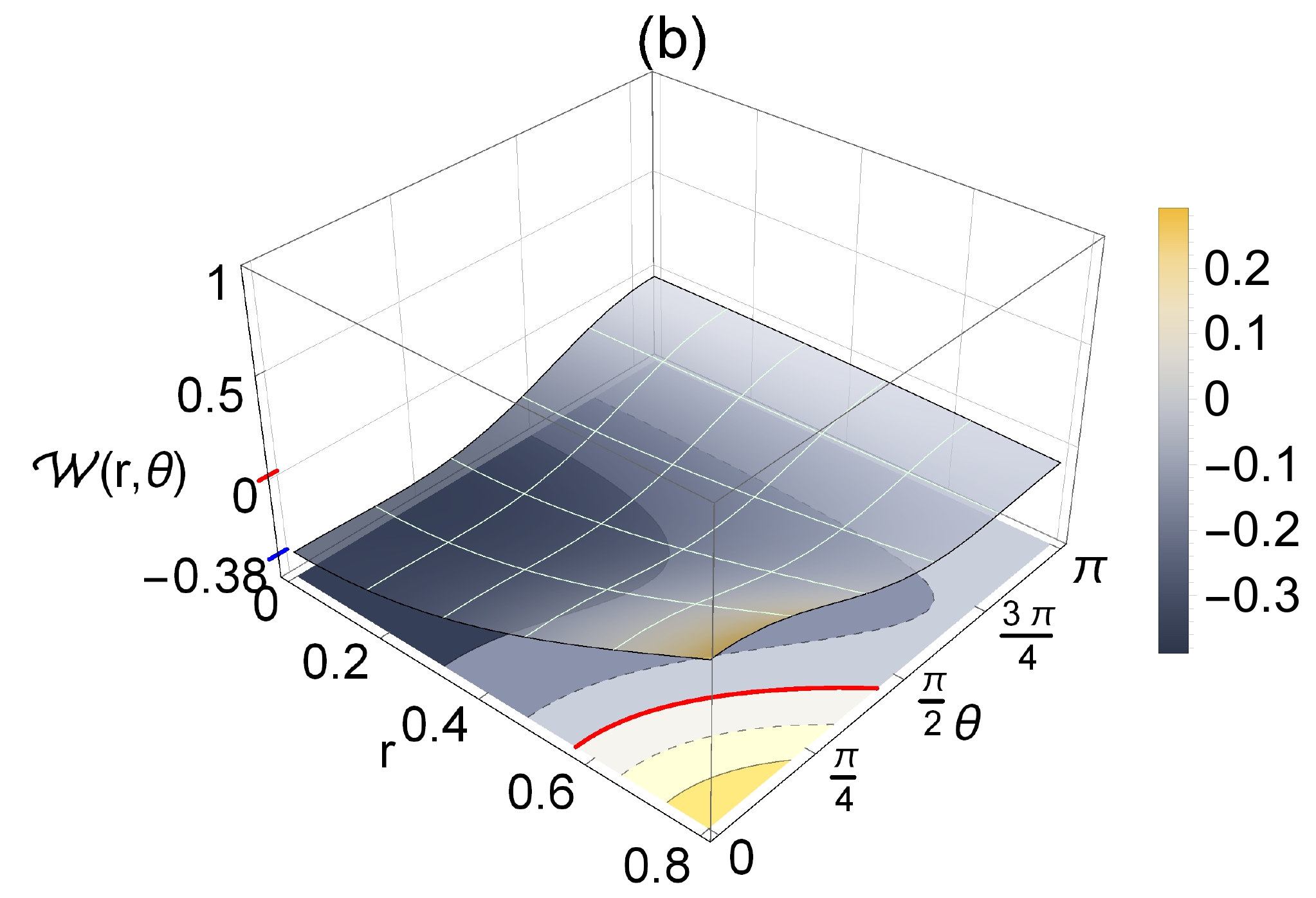}\\
 	\includegraphics[width=0.45\linewidth, height=5cm]{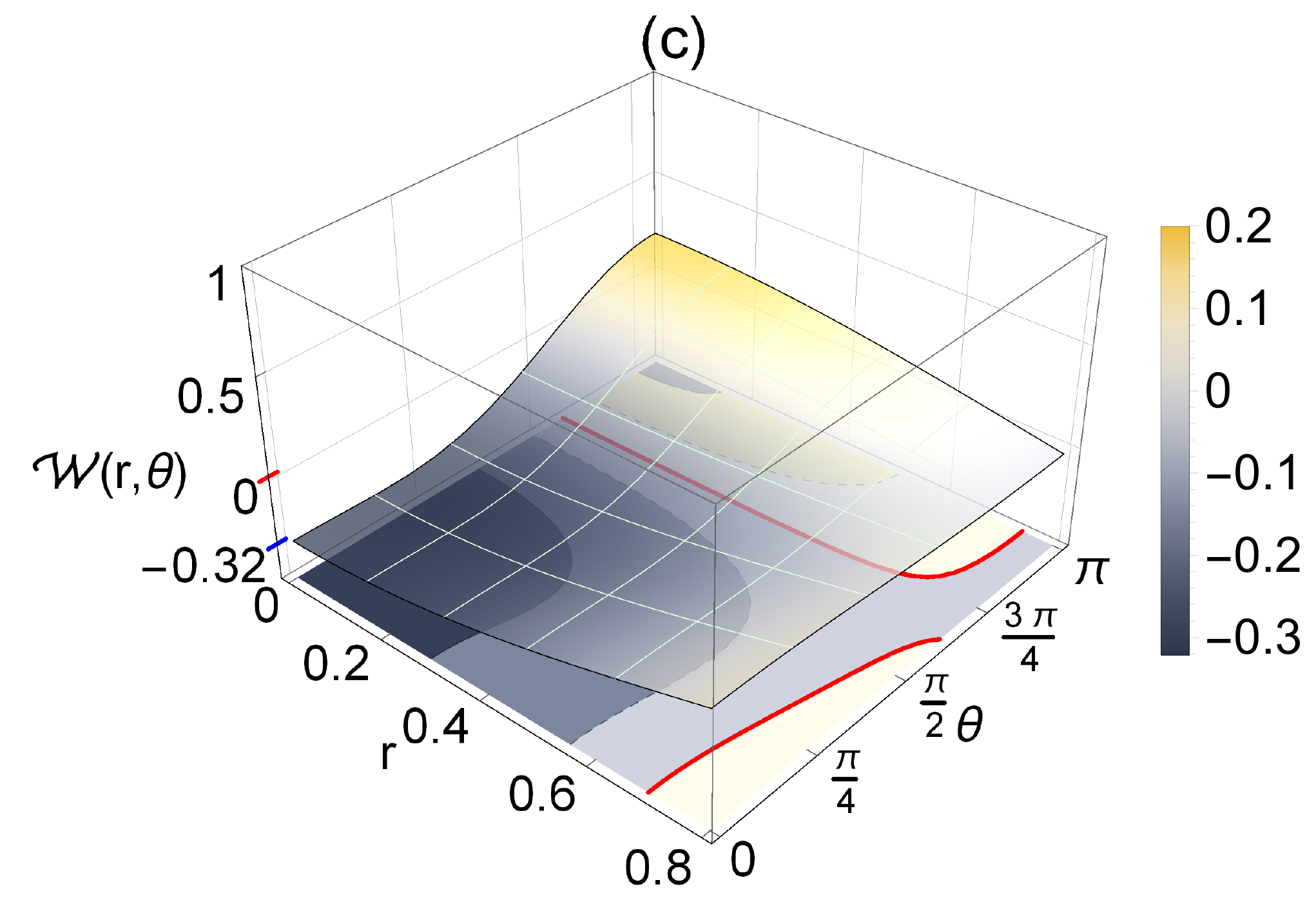}
 	\includegraphics[width=0.5\linewidth, height=5cm]{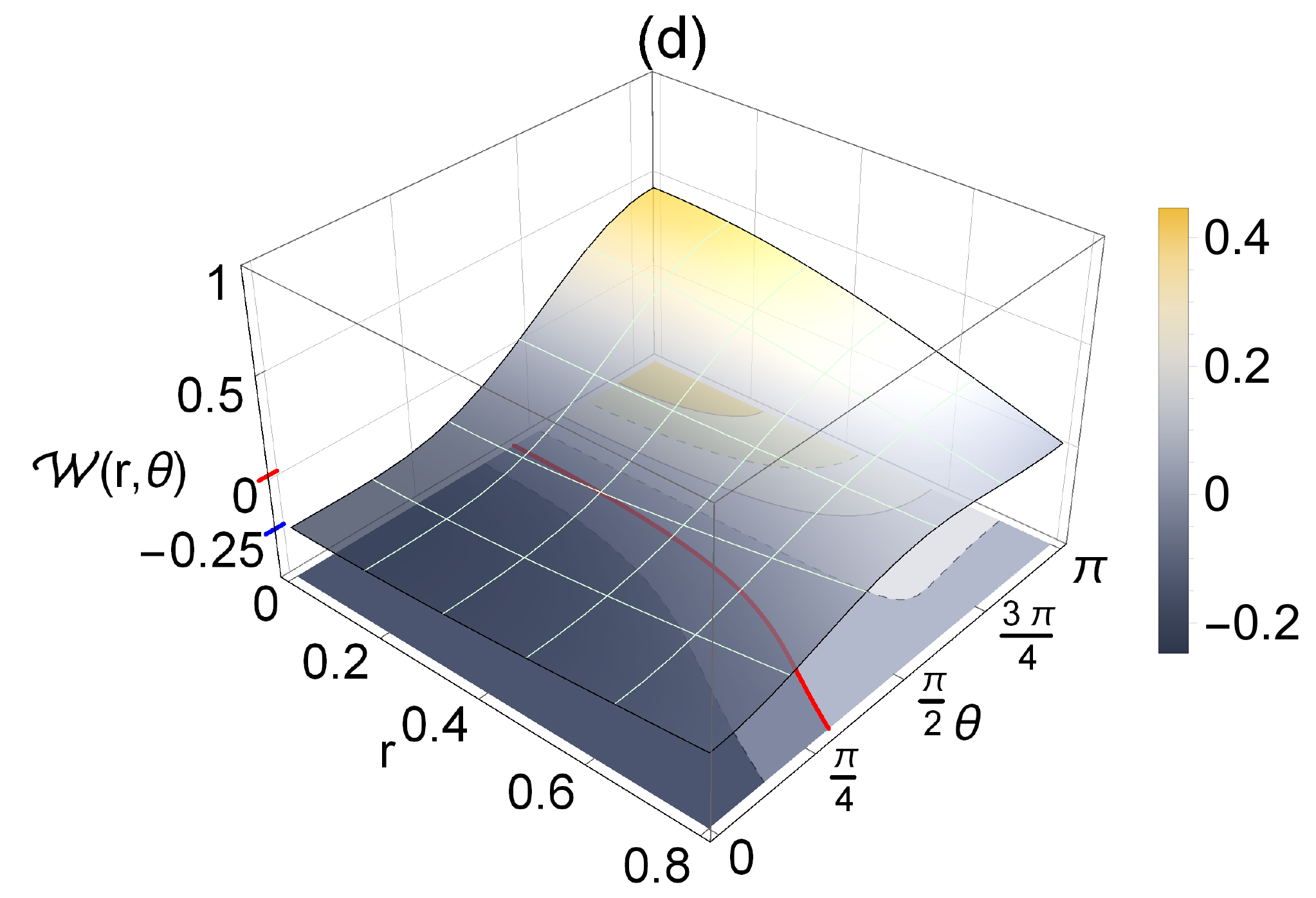}
 	\caption{ The effect of the amplitude damping channel on the Wigner function $W(r,~\theta)$, where  $ \phi=\pi $, and  (a) $p_{ad}=0.0$, (b) $p_{ad}=0.4$, (c) $p_{ad}=0.6$, and (d) $p_{ad}=0.8$.}
 	\label{fig:3.2}
 \end{figure}
 The initial density operator $ \hat{\rho}_{ab}(0) $ in Eq.(\ref{3.5}) evolves in the presence of the channel  $(\mathcal{C}_{ad})$ according to Eq.(\ref{3.6}) as follows:
 \begin{equation}\label{3.ad}
 \begin{split}
 \hat{\rho}^{ad}_{ab}&=\mathcal{B}_{11}|00\rangle \langle 00|+\mathcal{B}_{22}|10\rangle \langle 10|+ \mathcal{B}_{33}|01\rangle \langle 01|+\mathcal{B}_{44}|11\rangle \langle 11|
 +(\mathcal{B}_{14}|00\rangle\langle 11|+\mathcal{B}_{23}|10\rangle \langle 01|+ h.c.),
 \end{split}
 \end{equation}
 where

 \begin{equation}
 \begin{split}
&\mathcal{B}_{11}=cos^2 r \left[p_{ad} \varrho_{22}+ \left(p_{ad} \sin ^2 r+\cos^2 r\right)\varrho_{11} \right],\quad\mathcal{B}_{33}=(1-p_{ad}) \cos ^2r \left[\varrho _{22}+\varrho _{11} \sin ^2(r)\right],\\&
\mathcal{B}_{22}=(p_{ad} +\sin ^2 r \cos ^2 r+p_{ad} \sin ^4 r \big)\varrho _{11}+\big(2p_{ad}\sin ^2 r+ \cos ^2 r\big)\varrho _{22},\\& \mathcal{B}_{44}= (1-p_{ad}) \left((1+\sin ^4 r)\varrho _{11} +2\sin ^2r\varrho _{22}\right),\quad \mathcal{B}_{ij}=\sqrt{1-p_{ad}}  \varrho_{11} \cos ^2 r.
 \end{split}
 \end{equation}
 The Wigner function  in this case is given by,
 \begin{equation}\label{w.ad}
 \begin{split}
 W_{\hat{\rho}^{acc-ad}}(\theta,\phi)=&2\pi \big[\mathcal{B}_{11} \Psi^2_{11}+(\mathcal{B}_{22}+\mathcal{B}_{33})\Psi_{11}\Psi_{22} + \mathcal{B}_{44}\Psi^2_{22}+\mathcal{B}_{14} (\Psi^2_{12}+\Psi^2_{21})+2\mathcal{B}_{23}\Psi_{12}\Psi_{21}\big].
 \end{split}
 \end{equation}

  \begin{figure}[h!]
  	\centering
  	\includegraphics[width=0.45\linewidth, height=5cm]{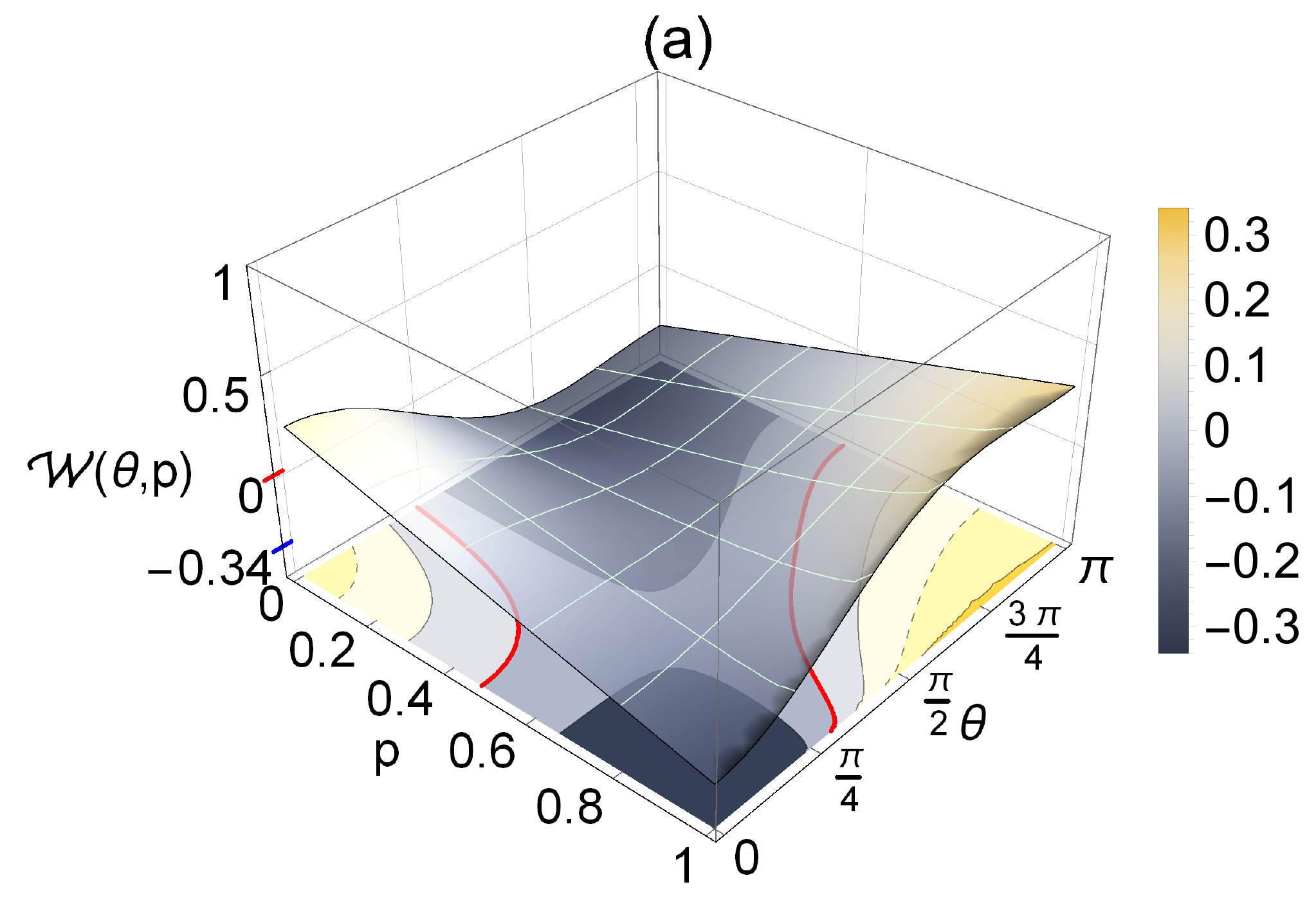}
  	\includegraphics[width=0.45\linewidth, height=5cm]{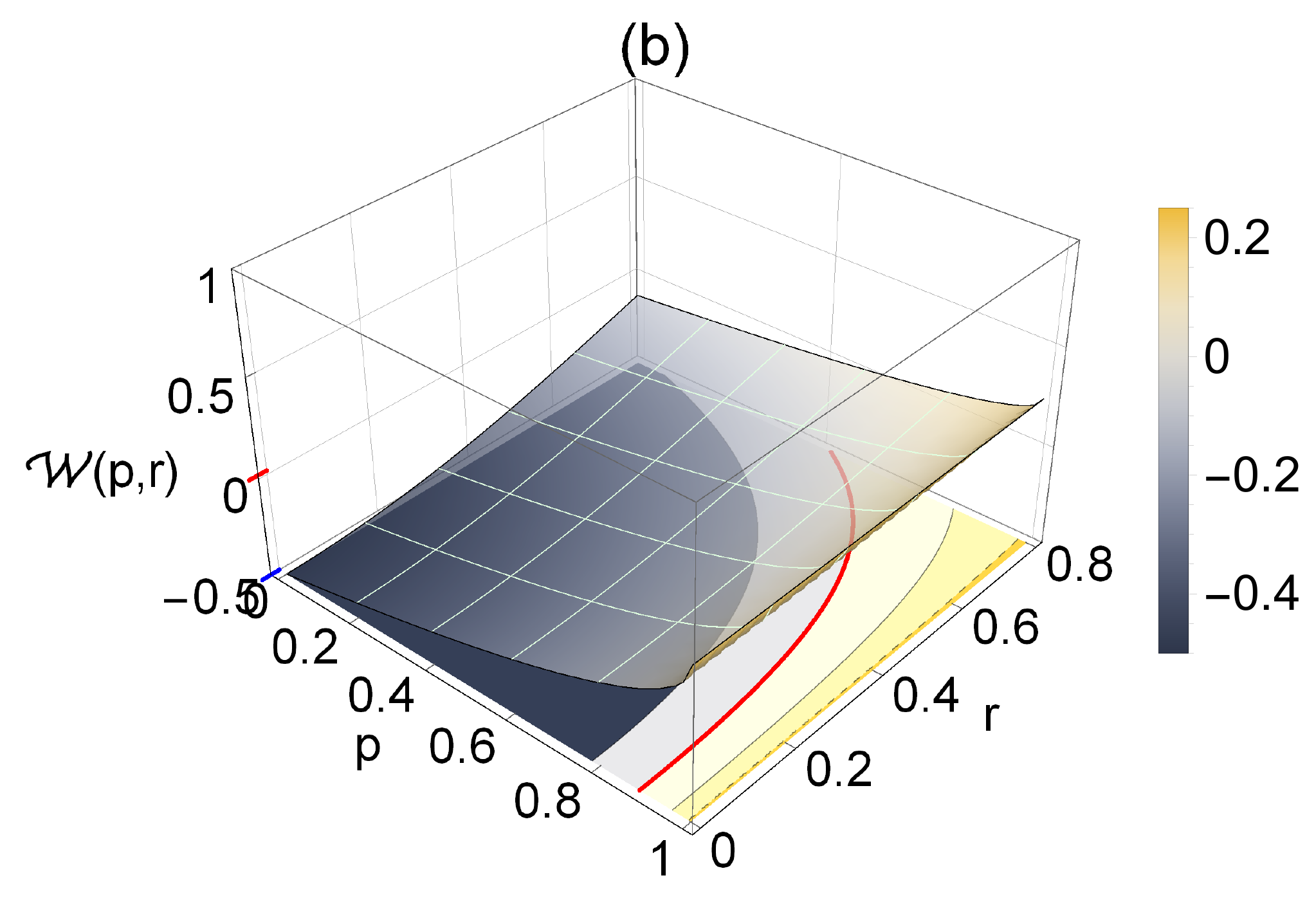}
  	\caption{ (a)  $W(p,\theta)$ at $ \phi=\pi $, $r=0.6$, and  (b) $W(p,r)$ at $\theta=\pi/2, \phi=\pi $.}
  	\label{fig:3.3}
  \end{figure}

\begin{figure}[h!]
	\centering
	\includegraphics[width=0.9\linewidth, height=5cm]{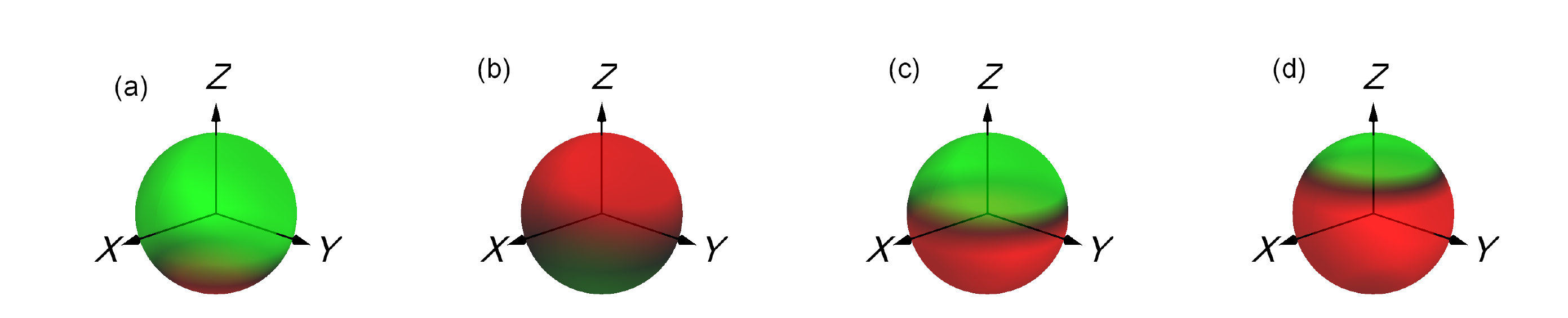}
	\caption{The behavior of $W(\theta,~\phi)$ on the Bloch sphere for the  accelerated system, where (a) $r=0.6,~ p_{ad}=0.6, (b) r=0.78,~ p_{ad}=0.6, (c)r=0.6,~ p_{ad}=0.8$, and  $(d)  r=0.78,~ p_{ad}=0.8.$}
	\label{fig:3.2.2}
\end{figure}

The effect of the amplitude damping channel is displayed in Fig.(\ref{fig:3.2}) at  different values of the channel strength $p_{ad}$. The behavior of $W(r,\theta)$ displays different effects of the channel strength, where at $p_{ad}=0$, the maximum entangled  state loses its quantum correlation at $r>0.5$ and $\theta<\pi/2$. Moreover, as it is  displayed in Fig.(\ref{fig:3.2}.a), the maximum value of the Wigner function is exhibited  as $r\to\infty$ and $\theta=0$. However, as one increases the channel' strength, the entangled behavior of the initial state is reproduced at larger acceleration. On the other hand, the minimum (maximum) values of the Wigner function are smaller than those displayed in Fig.(\ref{fig:3.2}.a). Different behaviors are displayed for   any value of $p_{ad}\in[0.5,~1]$, where the violation of inseparability is illustrated at large values of the parameter $\theta\in[3\pi/4,\pi]$ and any value of the acceleration. However, as one increases the channel strength ($p_{ad}=0.8$) the robustness of the inseparability is displayed at large acceleration and for any $\theta\in[0,\pi/4]$.

In Fig.(\ref{fig:3.3}.a),  Wigner function behavior  is displayed at a particular value of the acceleration, where we set $r=0.6$. The inseparability of the maximum accelerated entangled state is depicted  at  different intervals of the channel strength $p_{ad}$ and the  distribution angle $\theta$. The behavior of $W(\theta,~p)$  displays that the accelerated state keeps its inseparability at small values of $p<0.5$ and large values of $\theta\in[\pi/2]$ or large values of $p_{ad}\geq 0.5$ and small values of $\theta\leq\pi/4$.
The behavior of  $W(p,~r)$ at the particular value of $\theta=\pi/2$ and $\phi=\pi$ is displayed in Fig.(\ref{fig:3.3}.b). It is clear that, the accelerated state loses its separability gradually  as the acceleration increases, where the positive behavior of the Wigner function is performed at large large values of the acceleration parameter $r$ and  channel strength $p_{ad}>0.9$.

Fig.(\ref{fig:3.2.2}), displays the behavior of $W(\theta,~\phi)$ on the Bloch  sphere when the accelerated system passes through the amplitude damping  channel at some specific values of the channel strength $p_{ad}$ and the acceleration parameter $r$.  In Fig.(\ref{fig:3.2.2}.a,\ref{fig:3.2.2}.b), we fix the value of the channel strength to $p_{ad}=0.6$, while two values of the acceleration namely $r=0,6$, and $ 0.78$) are considered, respectively. It is clear that, the classical correlation appears as a small bottom cap in the region  $-1\leq z\leq -\frac{1}{2}$ . This means that, the channel strength transfers the quantum correlations which appear as upper cap on $0.5\leq z\leq 1$  to lower cap. Moreover, the area of the green surface is larger than that displayed in Fig.(\ref{fig:3.1.1}.b). However, as one increases the acceleration, the quantum and classical correlations are switched between the top and  the bottom of the sphere. Moreover, the  red area increases on the expense  of the green area $(-0.5\leq z\leq 1)$, namely the accelerated state  loses  its quantum correlations. In Figs.(\ref{fig:3.2.2}.c) and (\ref{fig:3.2.2}.d), we increase the value of channel strength to $p_{ad}=0.8$. It is clear that, the red area increases  and the  green area shrinks. However, by comparing Figs.(\ref{fig:3.2.2}.b) and (\ref{fig:3.2.2}.d), we can see that larger values of the channel strength re-create again a quantum correlations as an  upper cap in the region $0.5\leq z\leq 1$.

\subsection{ The bit-phase flip channel $\mathcal{C}_{bpf}$:}
The bit-phase flip  channel $(\mathcal{C}_{bpf}$) combines the error and the phase of the traveling density operator. This channel is described by the following  Kraus operators,
\begin{equation}
E_1=\sqrt{1-\frac{p_{bpf}}{2}} \ I_2 ,\qquad E_2=-i \sqrt{\frac{p_{bpf}}{2}}(|0\rangle \langle 1|- |1\rangle \langle 0|),
\end{equation}
Consequently,
\begin{equation}
\begin{split}
\hat{\rho}^{B-PF}_{ab}&=\mathcal{D}_{11}|00\rangle \langle 00|+\mathcal{D}_{22}|10\rangle \langle 10|+\mathcal{D}_{33}|01\rangle \langle 01|+\mathcal{D}_{44}|11\rangle \langle 11|+(\mathcal{D}_{14}|00\rangle\langle 11|+\mathcal{D}_{23}|10\rangle \langle 01|+h.c.),
\end{split}
\end{equation}
where,
\begin{equation*}
\begin{split}
&\mathcal{D}_{11}=\frac{1}{2} \cos ^2r \left[p_{bpf} \varrho _{22}+\nu^+\varrho _{11} \right],\quad \mathcal{D}_{22}=\frac{1}{2} \left[p_{bpf} \left(\varrho _{11}+\nu^+\varrho _{22}\right)+ 2 \tan ^2r \mathcal{D}_{11} \right],\\& \mathcal{D}_{33}=\frac{1}{2} \cos ^2r \left(\varrho _{11} \nu^- +(2-p_{bpf}) \varrho _{22}\right),\quad\mathcal{D}_{44}=\frac{1}{2} \left((2-p_{bpf}) \left(\varrho _{11}+\varrho _{22}\nu^-\right) + 2\tan ^2r \mathcal{D}_{33} \right),\\&  \mathcal{D}_{14}=\frac{-1}{2}  \cos ^2 r \left(p_{bpf} \varrho _{23}+(p_{bpf}-2) \varrho _{14}\right),
\ \ \mathcal{D}_{23}=\frac{-1}{2}  \cos ^2r \left(p_{bpf} \varrho _{23}+(p_{bpf}-2) \varrho _{14}\right) ,\\& \ \text{with} \ \nu^\pm=(1\pm(1-p_{bpf}) \cos 2r).
\end{split}
\end{equation*}
The Wigner function  for this system is thus given by,

\begin{equation}\label{wbpf}
\begin{split}
W^{(s)}_{\hat{\rho}^{B-PF-acc}}(\theta,\phi)=&2\pi \big[\mathcal{D}_{11} \Psi^2_{11}+(\mathcal{D}_{22}+\mathcal{D}_{33})\Psi_{11}\Psi_{22} + \mathcal{D}_{44}\Psi^2_{22}+\mathcal{D}_{14} (\Psi^2_{1,2}+\Psi^2_{21})+2\mathcal{D}_{23}\Psi_{12}\Psi_{21}\big].
\end{split}
\end{equation}

\begin{figure}[t!]
	\centering
	\includegraphics[width=0.45\linewidth, height=5cm]{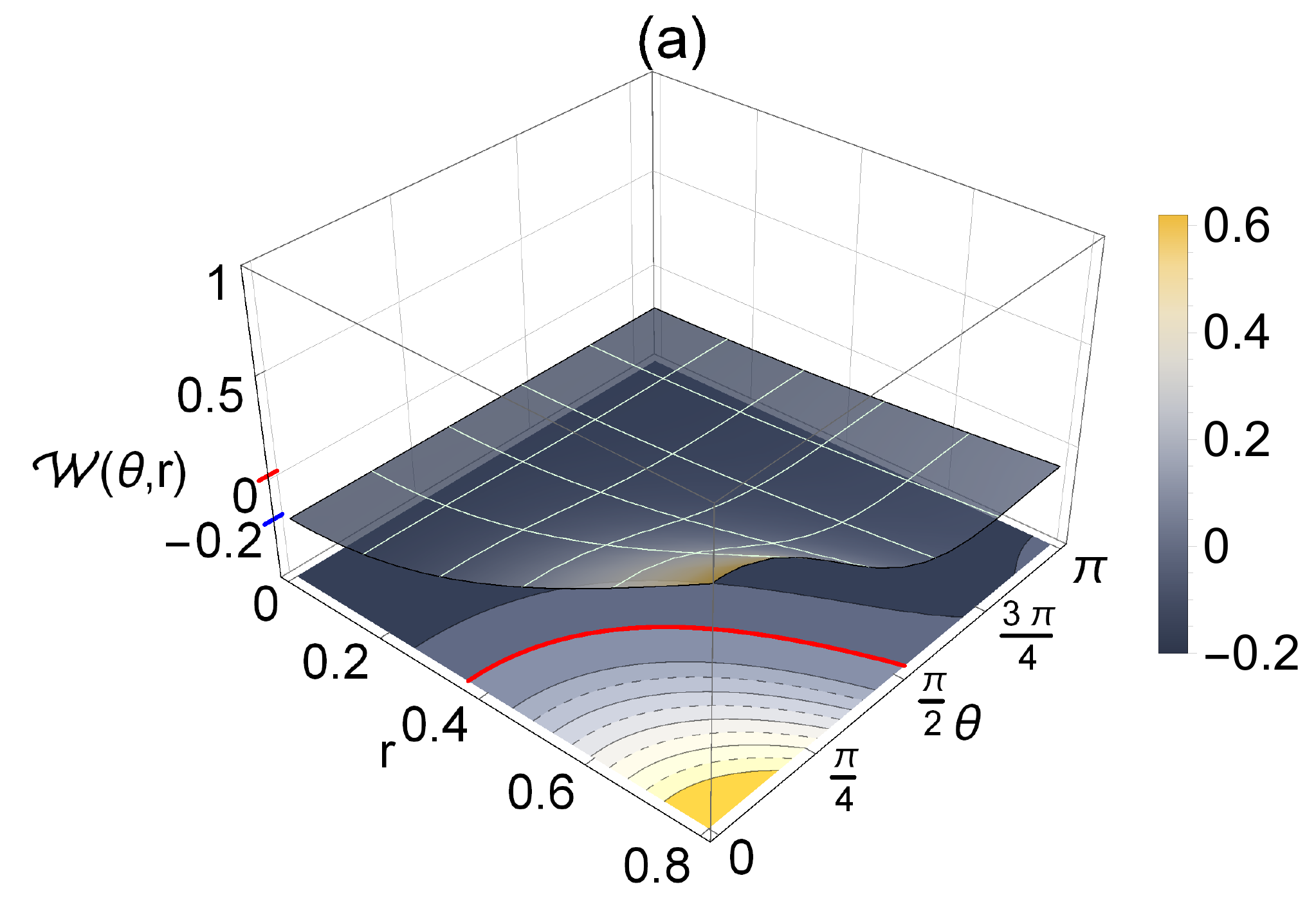}
	\includegraphics[width=0.45\linewidth, height=5cm]{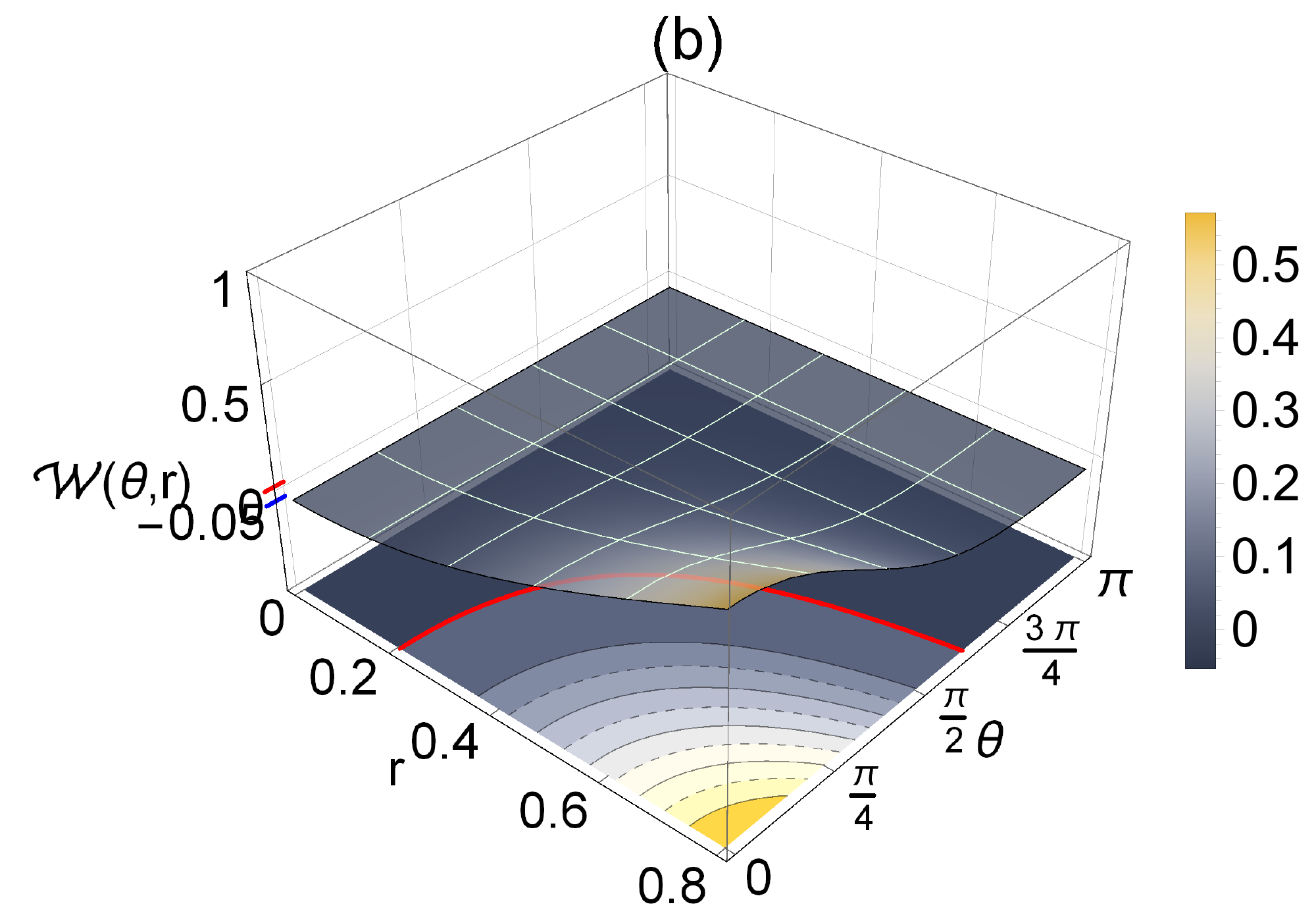}
	\includegraphics[width=0.45\linewidth, height=5cm]{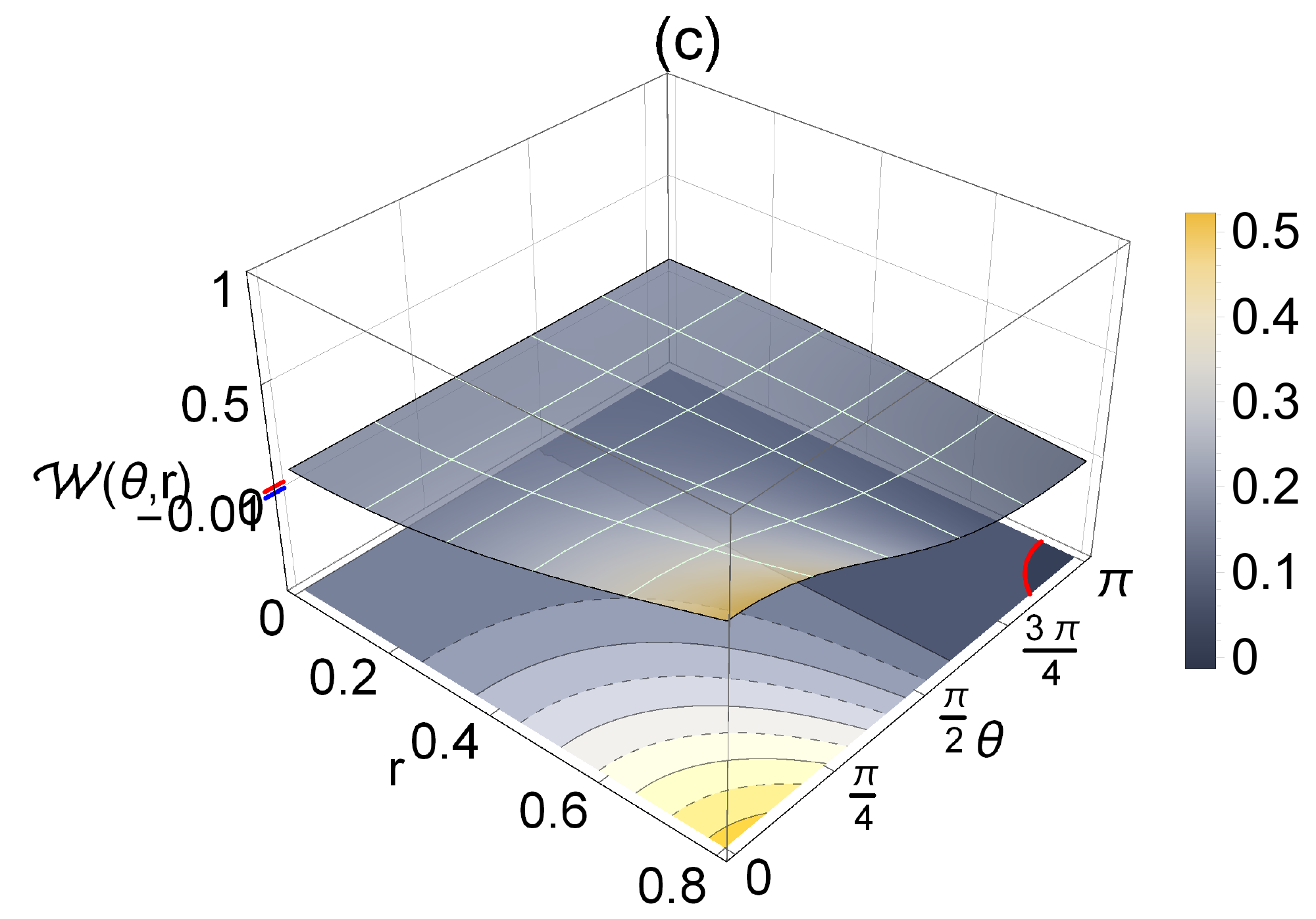}
	\caption{The same as Fig.(\ref{fig:3.1}) but the final accelerated state passes through the bit-phase flip channel $\mathcal{C}_{bpf}$.}
	\label{fig:3.4}
\end{figure}
\begin{figure}
	\centering
	\includegraphics[width=0.45\linewidth, height=5cm]{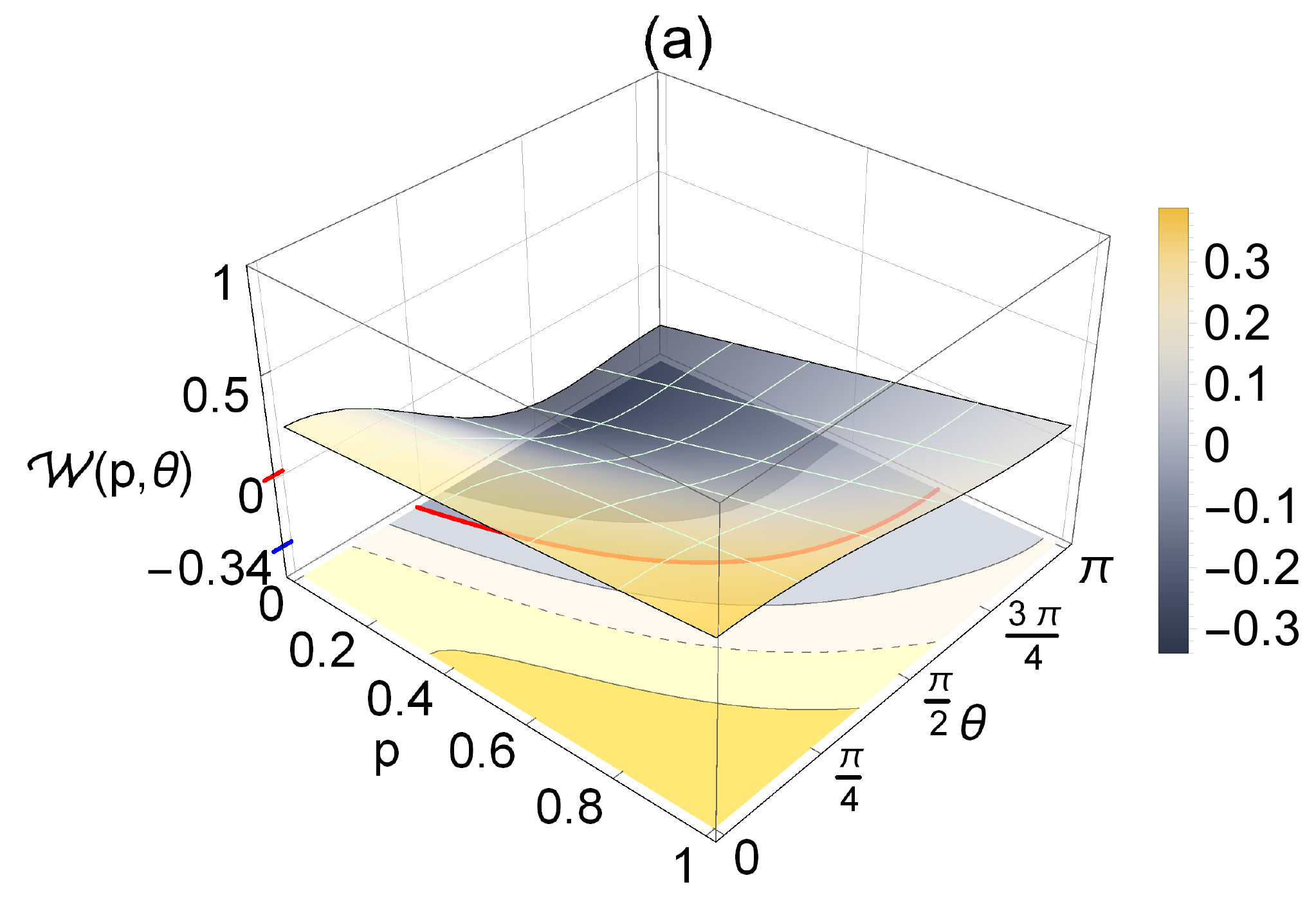}
	\includegraphics[width=0.45\linewidth, height=5cm]{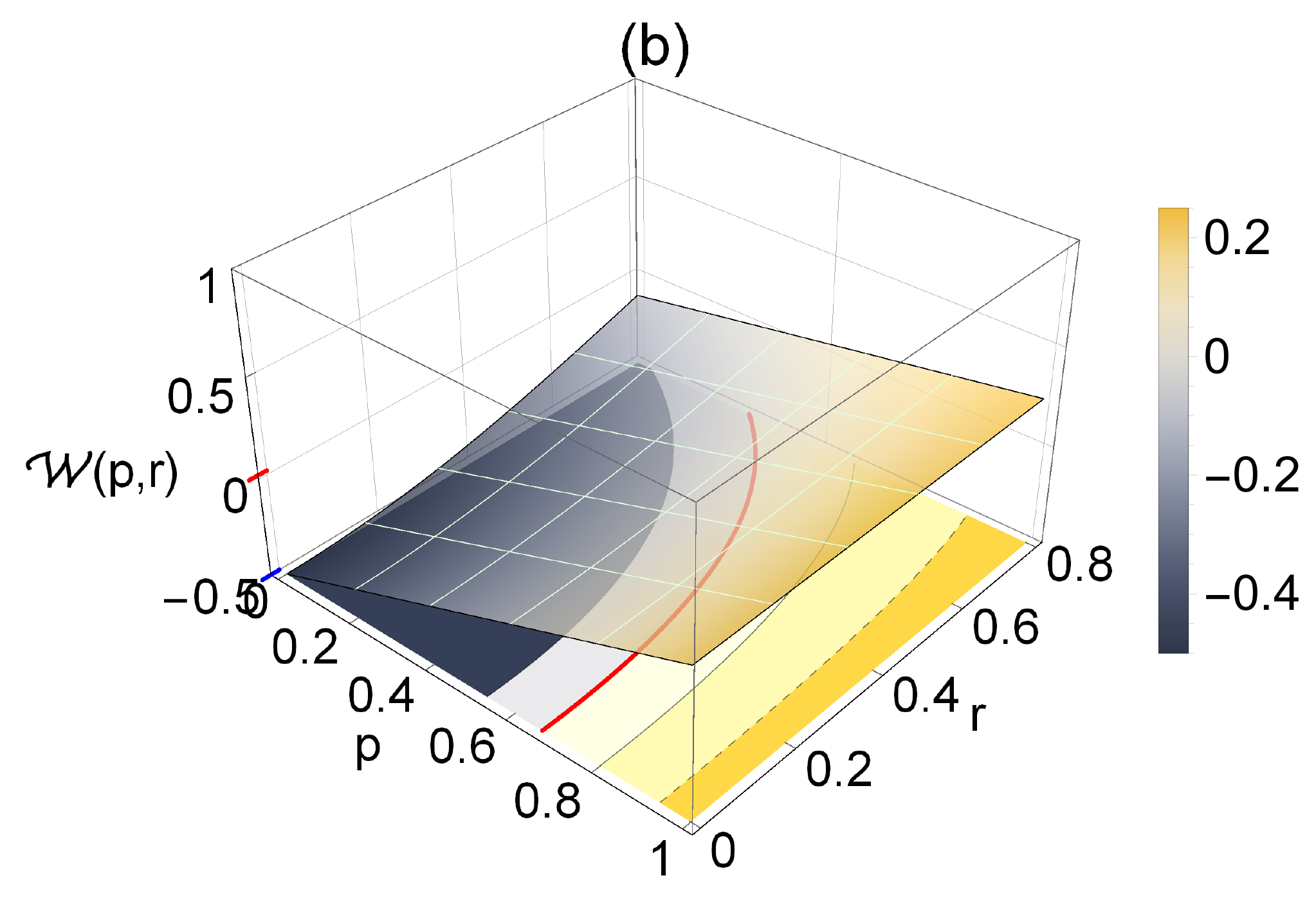}
	\caption{The same as Fig(\ref{3.3}), but the accelerated state passes through the phase flip channel $\mathcal{C}_{bpf}$}
	\label{fig:3.5}
\end{figure}

\begin{figure}[h!]
	\centering
	\includegraphics[width=0.85\linewidth, height=4.5cm]{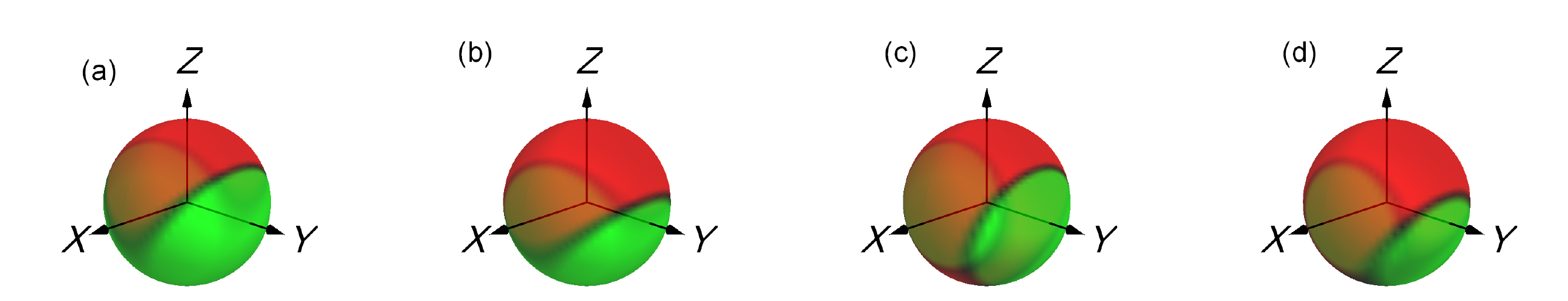}
	\caption{the same as fig(\ref{fig:3.2.2}) but the system influenced by bit phase flip channel.}
	\label{fig:3.3.3}
\end{figure}

The behavior of the Wigner function $W(r,\theta)$  at different values of the channel strength $p_{bpf}$ is displayed in Fig.(\ref{fig:3.4}). The behavior is similar to that displayed in Figs.(\ref{fig:3.2}b-\ref{fig:3.2}d), but the inseparability  of the accelerated state is displayed at smaller acceleration and larger values of the distribution angle  $\theta$.   For example,  as it is shown at $(p_{bpf}=0.4)$, the inseparability of the accelerated state is shown by Wigner function at $r<0.4$ and $\theta>\pi/2$. However, as one increases $p_{bpf}$, the negative behavior of the Wigner function is shown at smaller acceleration  and lager values of $\theta$ (see Fig.(\ref{fig:3.4}.b,\ref{fig:3.4}.c)). The minimum values  of the Wigner function $W(r,\theta)$ displayed for $\mathcal{C}_{bpf}$ are larger than those displayed for the amplitude damping channel $ \mathcal{C}_{ad}$. Meanwhile, the maximum values of $W(r,\theta)$ when the accelerated state passes through $\mathcal{C}_{bpf}$ are larger than those displayed when it passes through  $ \mathcal{C}_{ad}$. This means that, the  accelerated state which passes through the amplitude damping channel is more robust than that passes through the phase flip channel.

The effect of a particular   value of the acceleration $r$ on the behavior of the Winger function in the space ($p_{bpf}-\theta)$ is displayed in Fig.(\ref{fig:3.5}a). The inseparability of the accelerated state is depicted at small values of $p_{bpf}$ and larger values of $\theta>\pi/2$. However, in the space $(p_{bpf}-r)$, the behavior of   $W(r,\phi=\pi)$ is displayed at the specific value of $\theta=\pi/2$. The inseparability  behavior of the accelerated state is displayed at any value of the  acceleration $r$ and $p_{bpf}<0.5$.
From Figs.(\ref{fig:3.3}) and (\ref{fig:3.5}), one may conclude that, the accelerated state that passes through the amplitude damping channel $\mathcal{C}_{ad}$  the is more robust than that passes through the bit-phase flip channel $\mathcal{C}_{bpf}$.

Fig.(\ref{fig:3.3.3}) displays  Wigner function $W(\theta,~\phi)$ on a Bloch sphere for accelerated  system passes through the bit-phase channel at some specific values of the acceleration $(r)$ and the channel strength $p_{bpf}$. As it is illustrated in Fig.(\ref{fig:3.3.3}.a), the green area shrinks and appears in $0\leq y\leq 1$, which means that the  accelerated system loses some of its quantum correlation.  By comparing Figs.(\ref{fig:3.1.1}.b) and (\ref{fig:3.3.3}.a), the classical correlation  increases gradually  to cover the upper lune of the sphere. However, at large value of the acceleration parameter $r=0.78$, the green area shrinks more and the red area extends to cover most of the upper sphere.  At larger values of the channel strength $p_{bpf}$, the green area  decreases and appears as  a cap in the region  $0.5\leq y\leq 1$, and  consequently the amount of quantum correlation decreases.

 \subsection{The bit flip channel $\mathcal{C}_{bf}$:}
  The representation of Kraus operators $ E^a_i $ for this channel \cite{nielsen2002quantum} is given by,
\begin{equation}
E_1=\sqrt{1-\frac{p_{bf}}{2}} \ I_2 ,\qquad E_2=\sqrt{\frac{p_{bf}}{2}}(|0\rangle \langle 1|+ |1\rangle \langle 0|).
\end{equation}
\begin{figure}[H]
	\centering
	\includegraphics[width=0.45\linewidth, height=5cm]{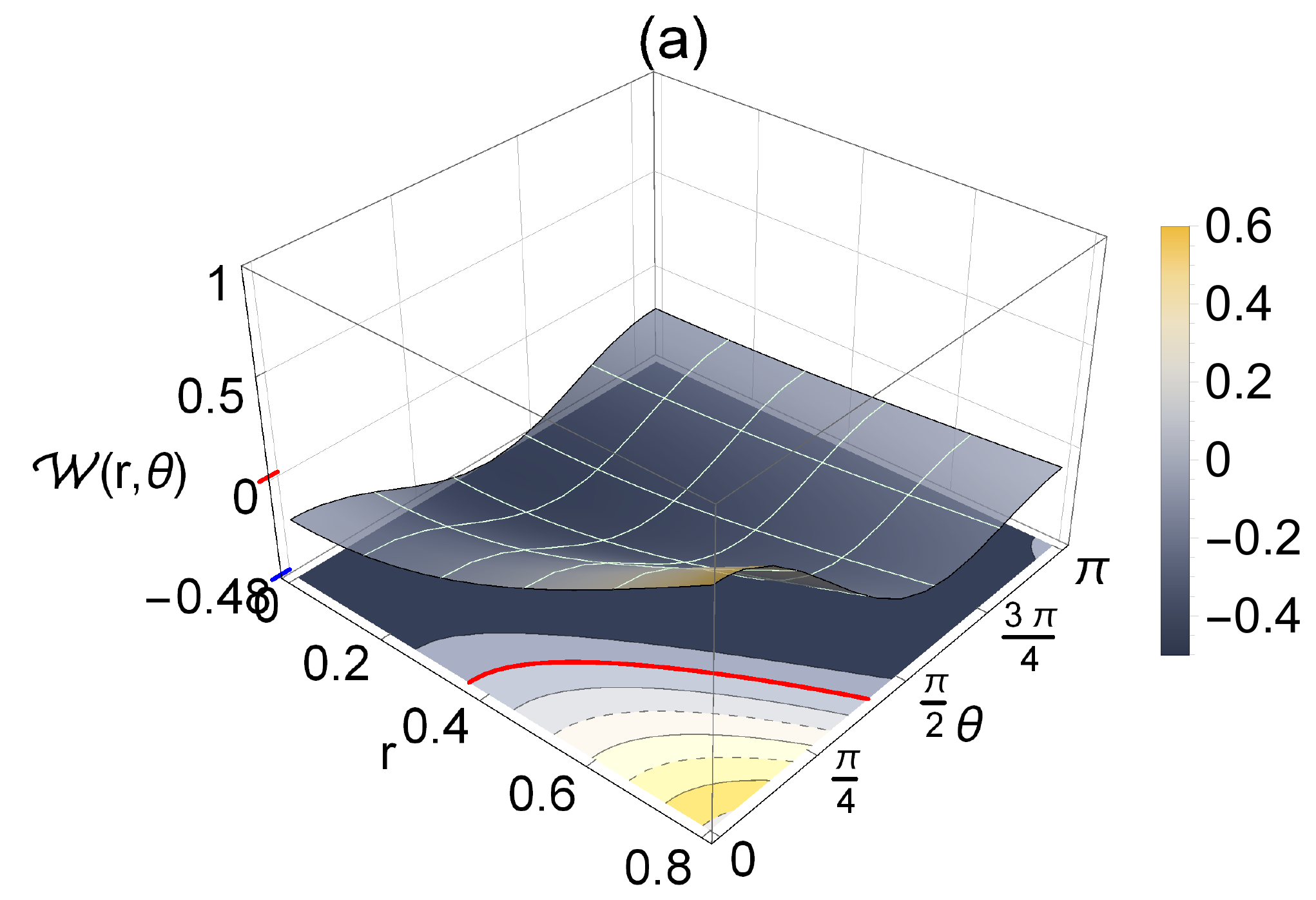}
	\includegraphics[width=0.45\linewidth, height=5cm]{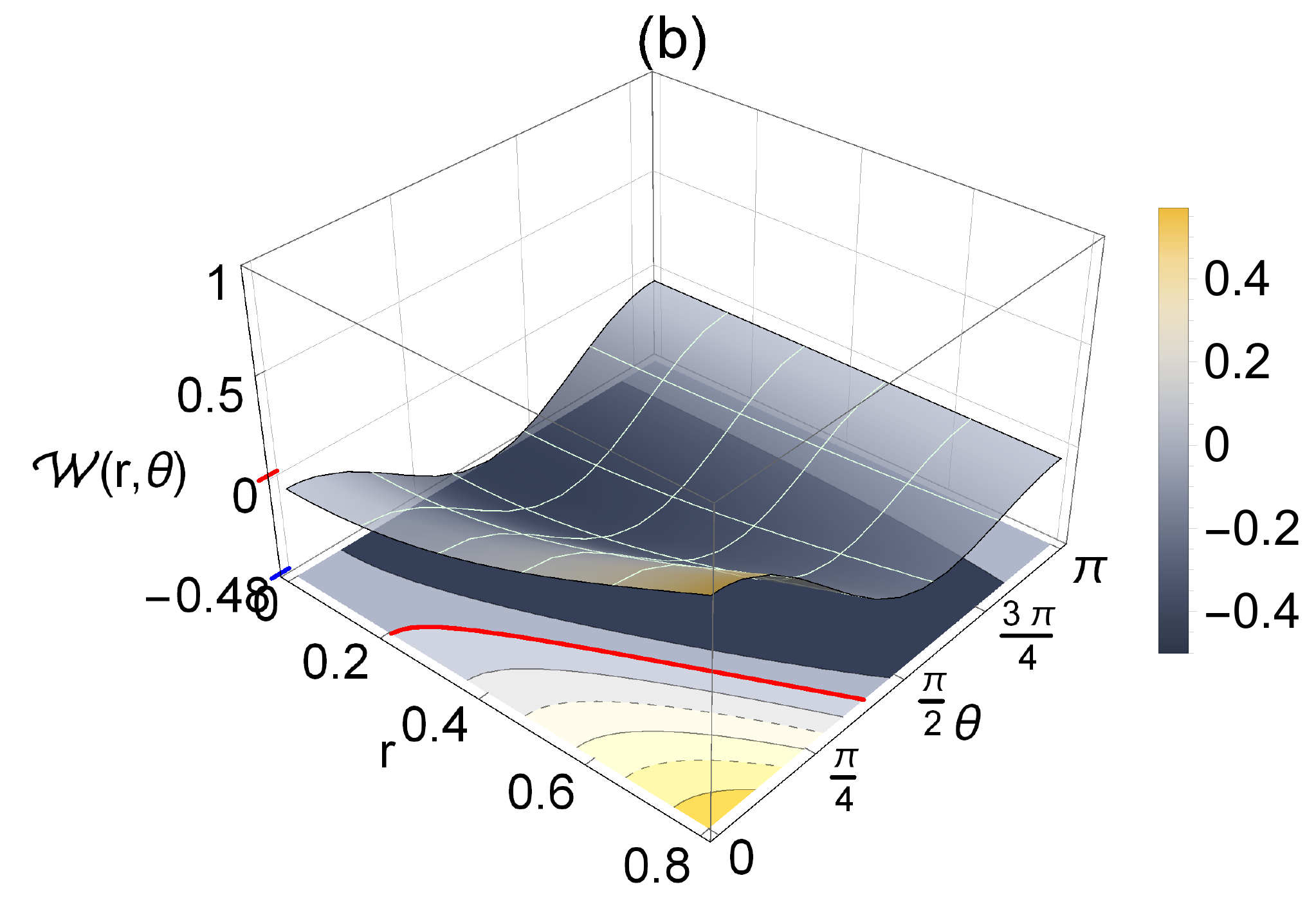}
	\includegraphics[width=0.45\linewidth, height=5cm]{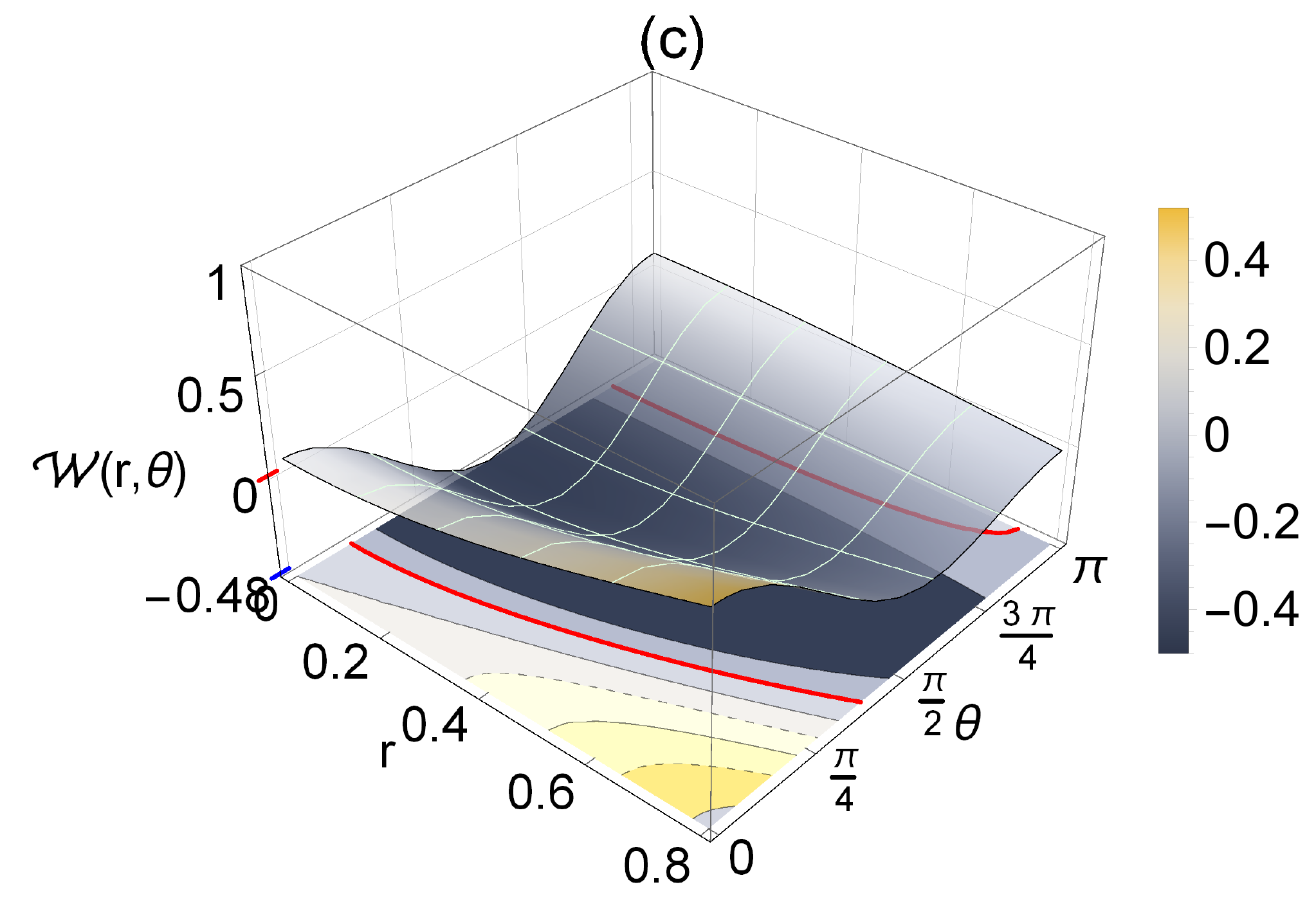}
	\caption{The same as Fig.(\ref{fig:3.2}), but the system passes through the bit flip channel $\mathcal{C}_{bf}$, where we set  $p_{bf}=0.4, 0.6$ and $0.8$ for (a), (b), and (c) respectively.}
	\label{fig:3.6}
\end{figure}
According to Eqs.(\ref{3.5}), (\ref{ch.3})and (\ref{3.6}), the final output state that passes through $\mathcal{C}_{bf}$ is given by,
\begin{equation}\label{bp}
\begin{split}
\hat{\rho}^{bf}_{ab}&=\mathcal{G}_{11}|00\rangle \langle 00|+\mathcal{G}_{22}|10\rangle \langle 10|+ \mathcal{G}_{33}|01\rangle \langle 01|+\mathcal{G}_{44}|11\rangle \langle 11|+ (\mathcal{G}_{14}|00\rangle\langle 11|+\mathcal{G}_{23}|10\rangle \langle 01|+h.c.),
\end{split}
\end{equation}
where,
$\mathcal{G}_{ii}=\mathcal{D}_{ii}
$ which is defined in the diagonal elements in the density operator in equation (\ref{bp}), while
\begin{equation*}                                                                        \mathcal{G}_{14}=\frac{1}{2}  \cos ^2 r \left(p_{bf} \varrho _{23}+(2-p_{bf}) \varrho _{14}\right),\ \
\mathcal{G}_{23}=\frac{1}{2}  \cos ^2r \left(p_{bf} \varrho _{23}+(2-p_{bf}) \varrho _{14}\right).
\end{equation*}

Subsequently, the Wigner function is obtained from Eq.(\ref{3.4}) as:

\begin{equation}
\begin{split}
W_{\hat{\rho}^{BF}}(\theta,\phi)=&2\pi \big[\mathcal{G}_{11} \Psi^2_{11}+(\mathcal{G}_{22}+\mathcal{G}_{33})\Psi_{11}\Psi_{22} +\mathcal{G}_{44}\Psi^2_{22}+\mathcal{G}_{14} (\Psi^2_{12}+\Psi^2_{2,1})+2\mathcal{G}_{23}\Psi_{1,2}\Psi_{21}\big].
\end{split}
\end{equation}

\begin{figure}[h!]
	\centering
	\includegraphics[width=0.45\linewidth, height=5cm]{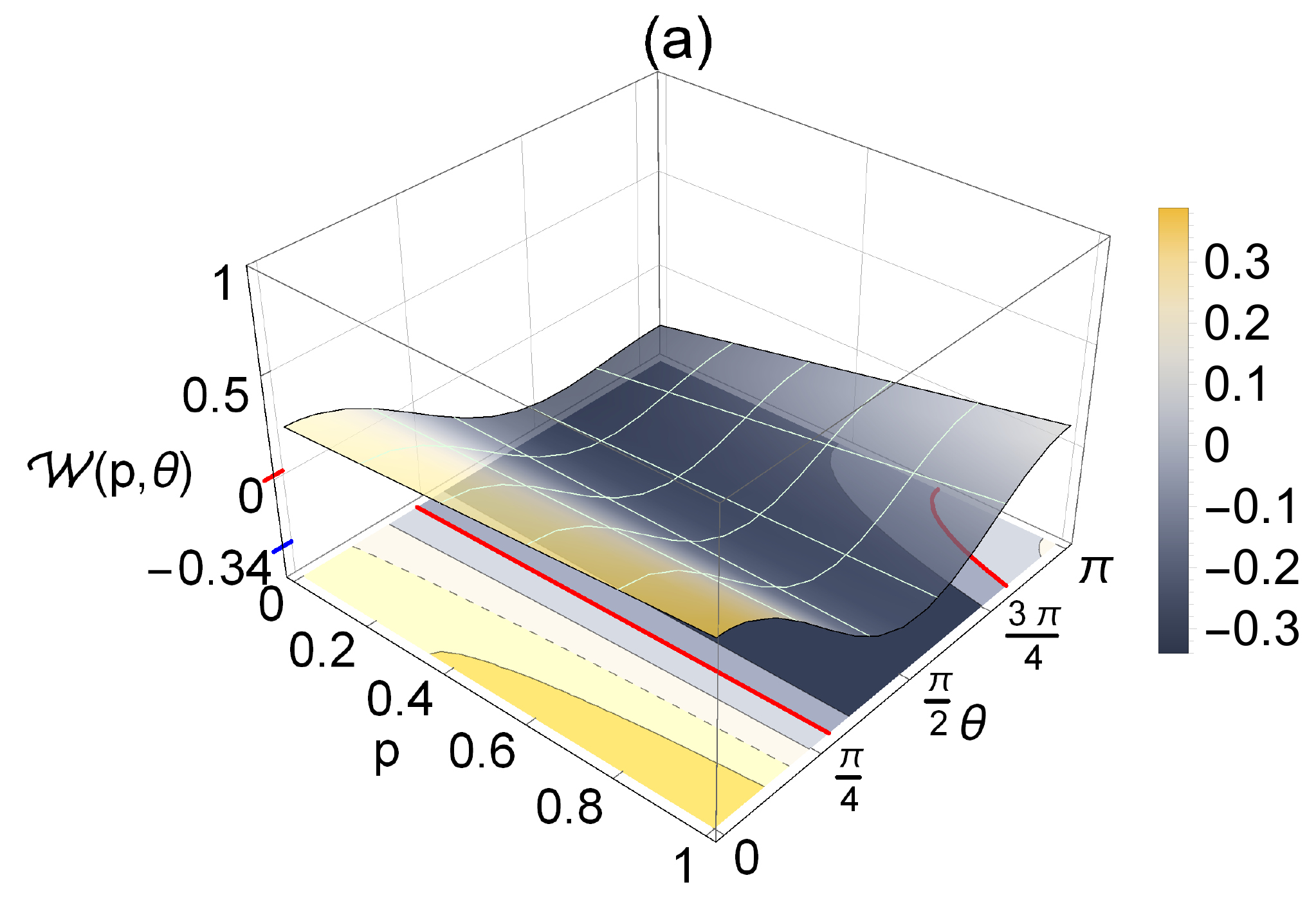}
	\includegraphics[width=0.45\linewidth, height=5cm]{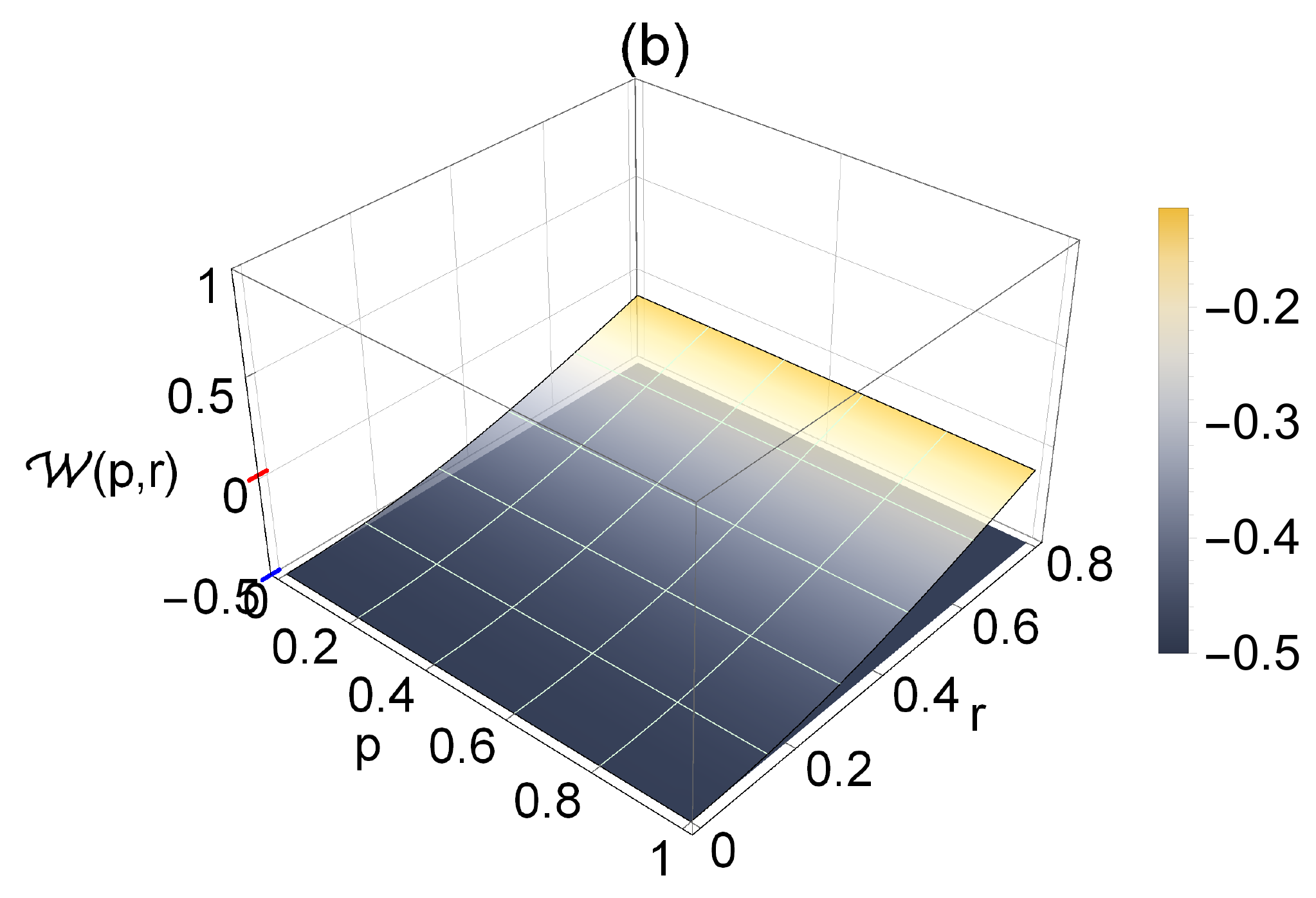}
	\caption{ The same as Fig.(\ref{fig:3.3}), but the accelerated state passes through the $\mathcal{C}_{bf}$.}
	\label{fig:3.7}
\end{figure}

\begin{figure}[h!]
	\centering
	\includegraphics[width=0.85\linewidth, height=4.5cm]{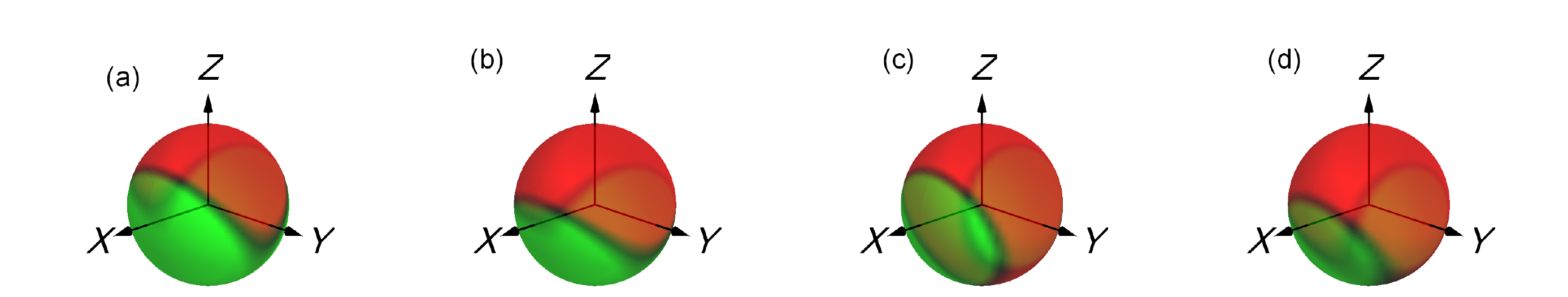}
	\caption{the same as fig(\ref{fig:3.2.2}) but the system influenced by bit Flip channel.}
	\label{fig:3.3.4}
\end{figure}

Fig.(\ref{fig:3.6}), displays the behavior of $W(r,~\theta)$ at different values of the channel strength. The general behavior shows that at large acceleration, the accelerated state loses its quantum correlation as the channel strength increases. The distribution  parameter $\theta$, may play as a control parameter to keep the survival of the quantum correlation of the accelerated state.  Moreover, at small values of the channel strength ($p_{bf}=0.3)$, the inseparability behavior (quantum correlation) is performed at small acceleration $r<0.4$ and $\theta<\pi/4$.  On the other hand, at large values of the channel strength $(p_{bf}=0.8)$, the inseparability of the accelerated state  is displayed in the interval $\pi/2<\theta<3\pi/7$.

The behavior of the Wigner function at some specific values of $ r$ and $\theta$ is shown in Figs.(\ref{fig:3.7}.a),(\ref{fig:3.7}.b), respectively. It is clear  from Fig.(\ref{fig:3.7}.a), the negative behavior of $W(p_{bf}~,\theta)$ is independent of the channel strength, if  one chooses $\theta\in[\pi/4~,3\pi/4]$. However, if we set $\theta=\pi/2$ and $\phi=\pi$, the  behavior of quantum correlation of the accelerated state is depicted  at values of the channel strength and the acceleration parameter in the plan $(p_{bf}-r)$ (see Fig.(\ref{fig:3.7}.b)).

Fig.(\ref{fig:3.3.4}) shows that, on the Bloch sphere, the behavior of $W(\theta~,\phi$) is almost similar to that displayed for the phase-bit flip channel (see Fig.\ref{fig:3.6}). However, the sectors that predicted the red and green areas are   rotated. However, the green area (quantum correlation) is displayed in the region $0\leq x\leq 1$, while the red area  (classical correlations) are displayed in the region $-1\leq x\leq 0$.

\subsection{Phase Flip  Channel, $\mathcal{C}_{pf}$}

Physically the channel $\mathcal{C}_{pf}$ illustrates any change on the phase that can occur on the transported state between Alice and Bob. The set of Kraus operators $ E^a_i $ corresponding to $\mathcal{C}_{pf}$ channel are defined as:

\begin{equation}
E_1=\sqrt{1-\frac{p_{pf}}{2}} \ I_2 ,\qquad E_2=\sqrt{\frac{p_{pf}}{2}}(|1\rangle \langle 1|- |0\rangle \langle 0|),
\end{equation}

Consequently, the final output state is given by,
\begin{equation}
\begin{split}
\hat{\rho}^{pf}_{ab}&= \mathcal{A}_{11} |00\rangle \langle 00| + \mathcal{A}_{22} (|01\rangle \langle 01|+|10\rangle \langle 10|)+\mathcal{A}_{33} |11\rangle \langle 11| +  (1-p_{pf})(\mathcal{A}_{14} |00\rangle \langle 11|+\mathcal{A}_{23} |10\rangle \langle 01|+h.c.).
\end{split}
\end{equation}
For this case, the Wigner function takes the form,
\begin{equation}\label{wpf}
\begin{split}
W_{\hat{\rho}^{pf}}(\theta,\phi)=&2\pi \big[\mathcal{A}_{11} \Psi^2_{11}+\mathcal{A}_{33}\Psi^2_{22}+2\mathcal{A}_{22}\Psi_{11}\Psi_{22}  + (1-p_{pf})(\mathcal{A}_{14} (\Psi^2_{12}+\Psi^2_{21})+2\mathcal{A}_{23}\Psi_{12}\Psi_{21})\big].
\end{split}
\end{equation}
\begin{figure}[H]
	\centering
	\includegraphics[width=0.45\linewidth, height=5cm]{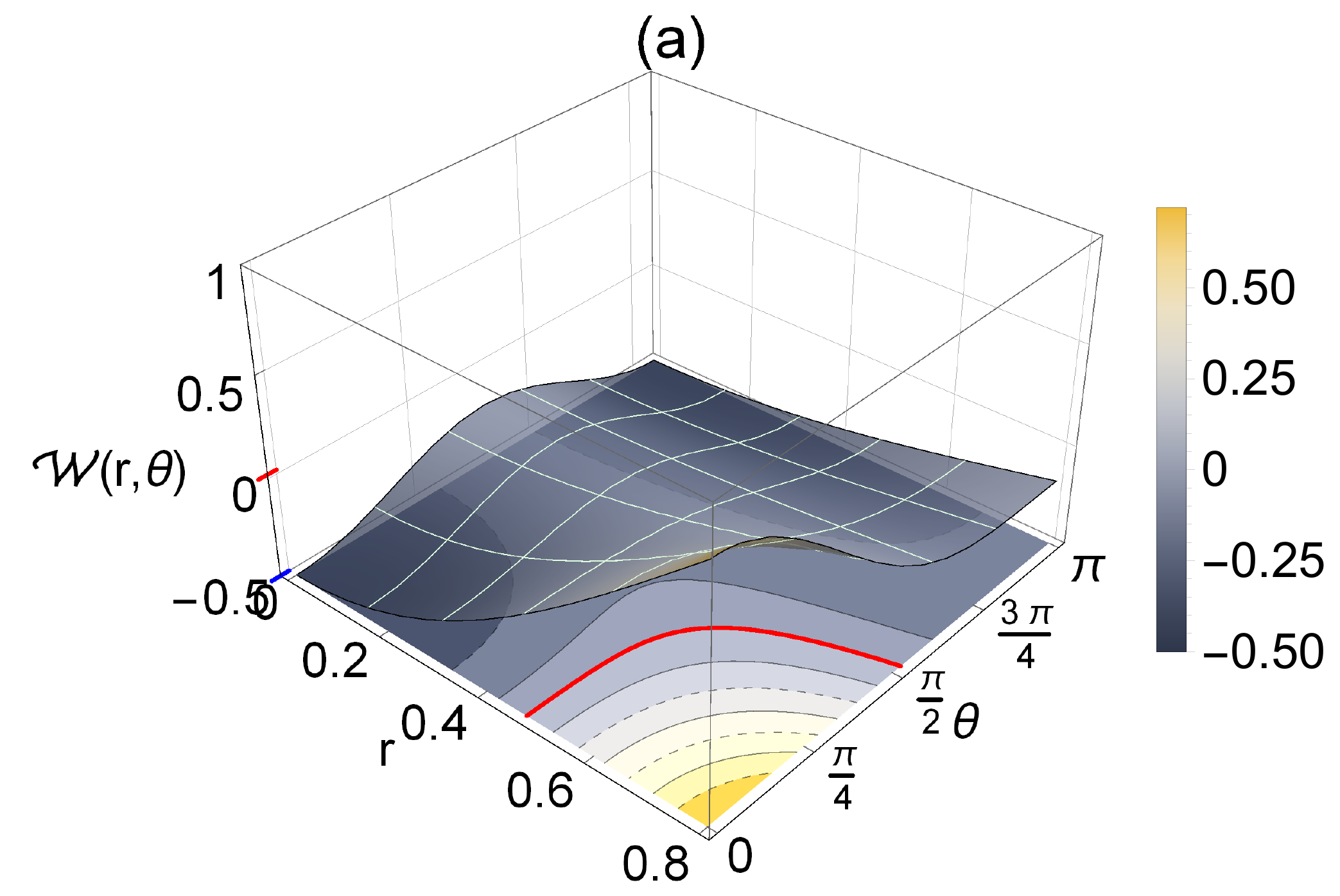}
	\includegraphics[width=0.45\linewidth, height=5cm]{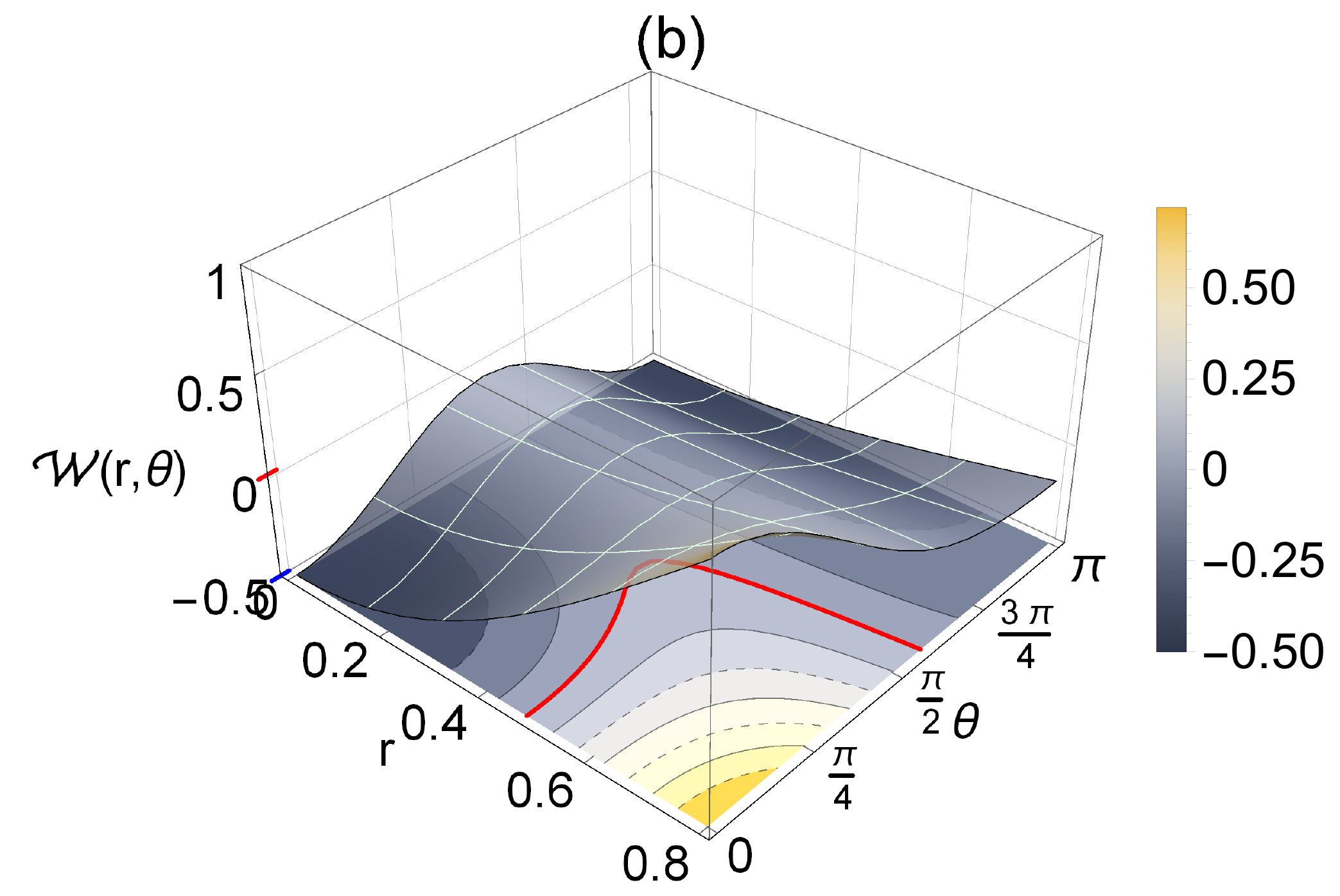}
	\includegraphics[width=0.45\linewidth, height=5cm]{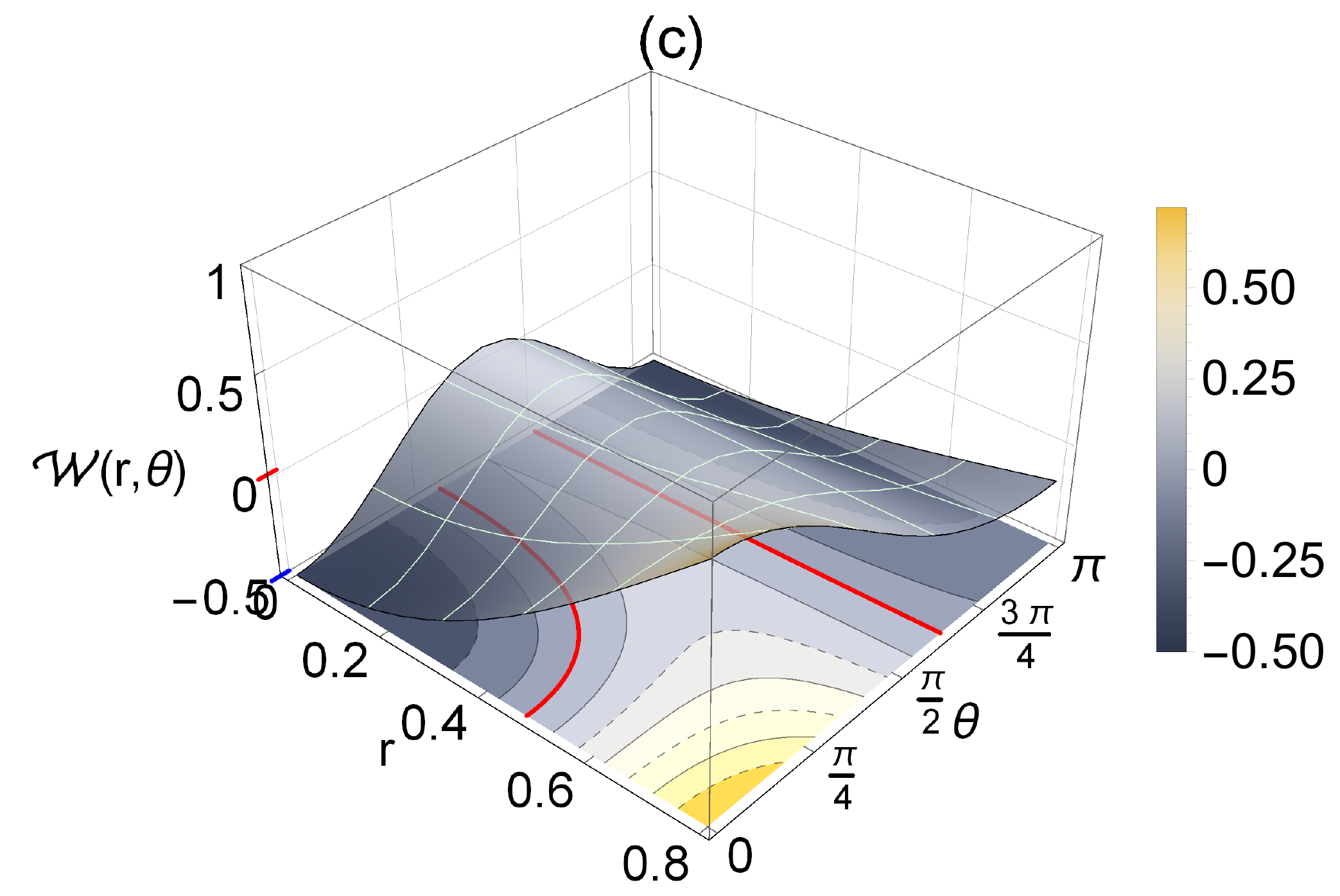}
	\caption{The same as Fig.(\ref{fig:3.1}) but the final accelerated state passes through the phase flip channel $\mathcal{C}_{pf}$.}
	\label{fig:3.8}
\end{figure}
\begin{figure}
	\centering
	\includegraphics[width=0.45\linewidth, height=5cm]{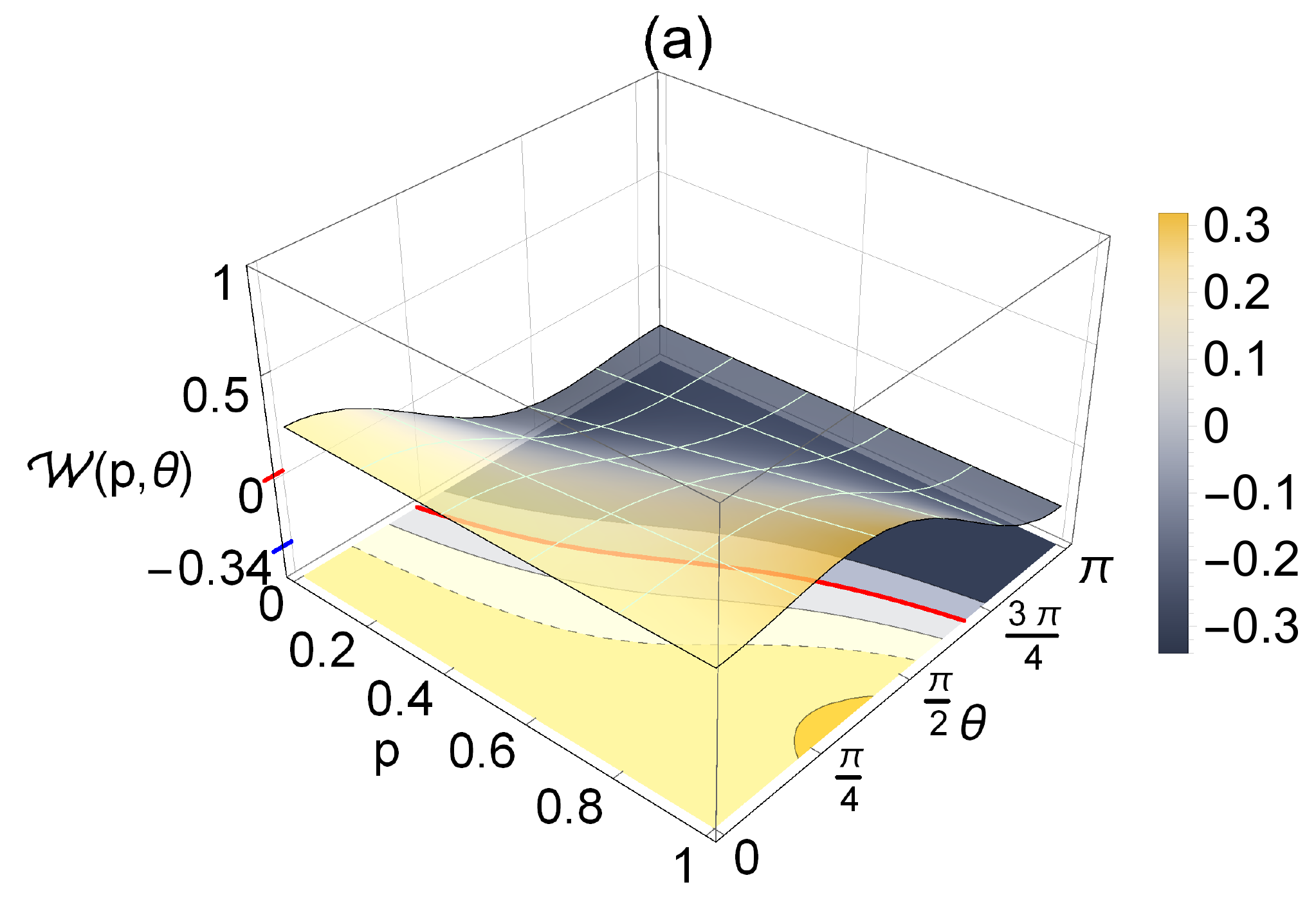}
	\includegraphics[width=0.45\linewidth, height=5cm]{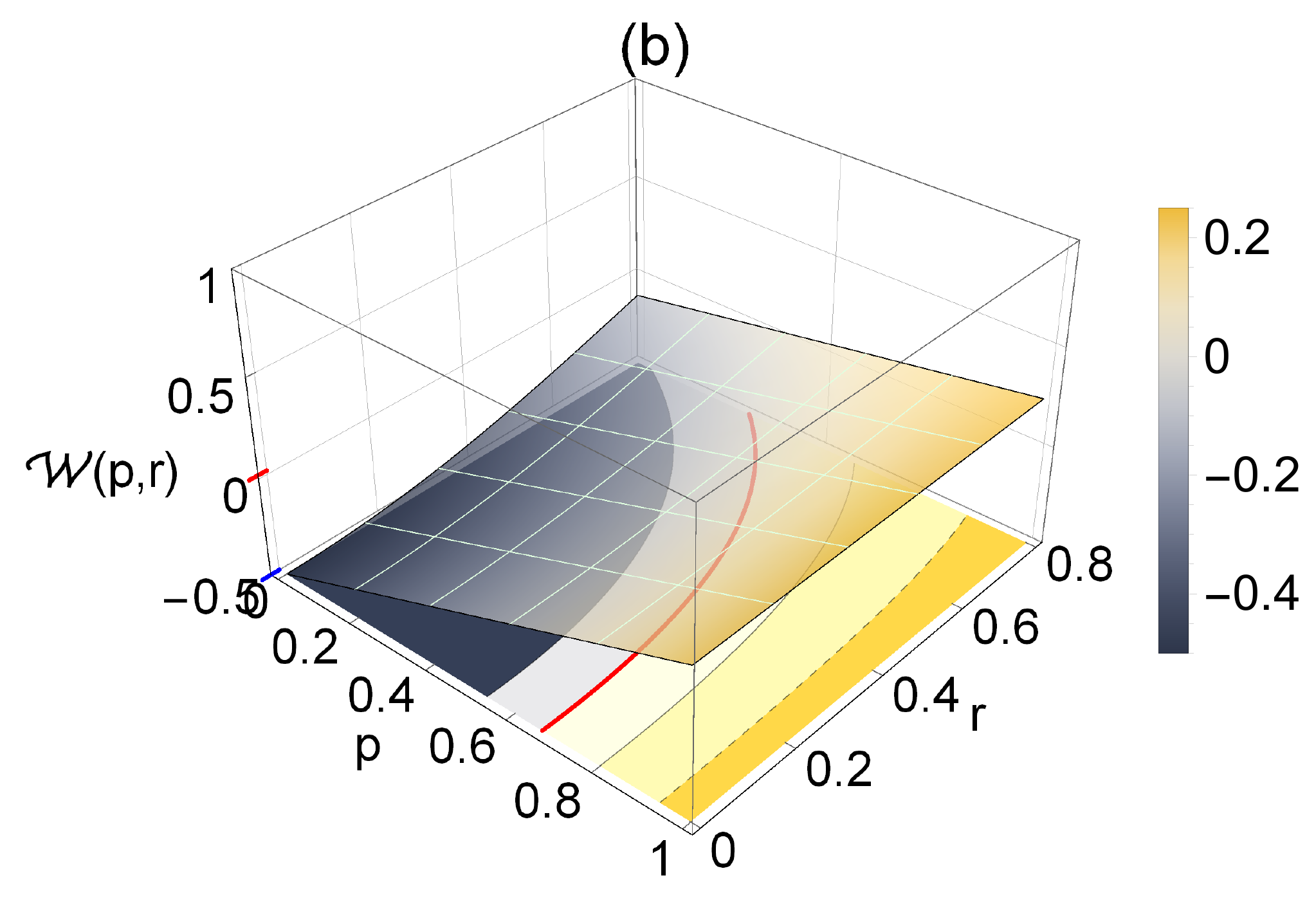}
	\caption{The same as Fig(\ref{fig:3.3}), but the accelerated state passes through the phase flip channel $\mathcal{C}_{pf}$.}
	\label{fig:3.9}
\end{figure}
\begin{figure}[t!]
	\centering
	\includegraphics[width=0.85\linewidth, height=4.0cm]{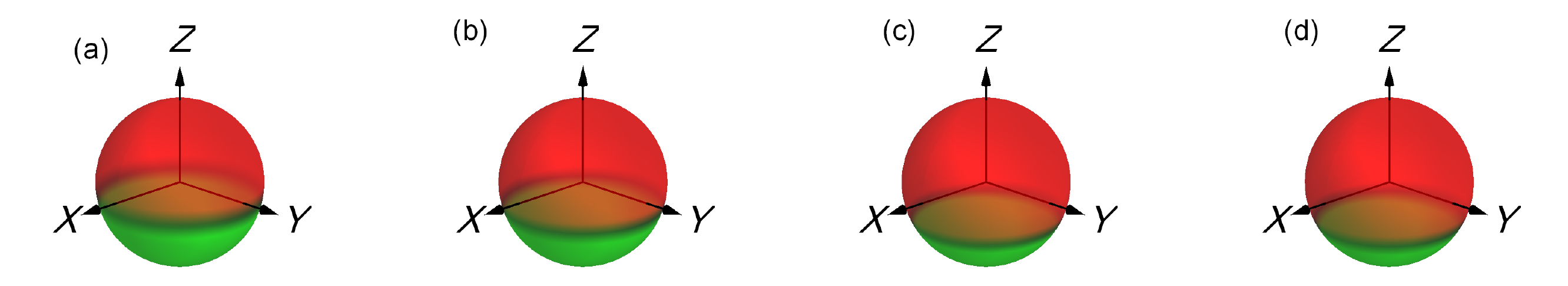}
	\caption{the same as fig(\ref{fig:3.2.2}) but the system influenced by phase flip channel.}
	\label{fig:3.3.5}
\end{figure}
The effect of the phase flip channel on the behavior of the Wigner function  in the plane $(r,~\theta)$ at some specific values of the channel strength $p_{pf}$ is displayed in Fig.(\ref{fig:3.8}).  Similarly, the negative behavior of the Wigner function $W(r,~\theta)$ in the plane $(r,~\theta)$ depends on different intervals of the acceleration and the distribution angle $\theta$. In general, the minimum (maximum) bounds of the Wigner function do not depend on the values of the channel strength. It is clear that, the inseparability  of the acceleration state is independent of the acceleration at  larger values of the distribution  angle $\theta>\pi/2$. The negativity of  $W(r,~\theta)$ and consequently the quantum correlation is performed at small  acceleration $r<0.5$ and small values of $\theta<\pi/4$. Moreover, at large value of the channel strength $(p_{pf}=0.8)$, the classical correlation is displayed at value of $r$ and $\pi/4<\theta<\pi/2$.

In Fig.(\ref{fig:3.9}.a), the behavior of the Wigner function in the plane $(p_{bf},\theta)$ is displayed at $r=0.6$. It is clear that, $W(p,~\theta)$ decreases gradually as one increases both parameters $p_{pf}$ and $\theta$. The minimum negative value of the Wigner function is depicted at any value of the strength parameter $p_{pf}$ and $\theta>3\pi/4$.  In the plan $(p_{bf}-r)$, the behavior of   $W(p,~r)$ at $\theta=\pi/2$ and $\phi=\pi$, as it is displayed in Fig.(\ref{fig:3.9}.b),  is similar to that  displayed for  $\mathcal{C}_{ad}$, $\mathcal{C}_{fp}$ and $\mathcal{C}_{pf}$. However, only the intervals of these parameters, at which the inseparability of the accelerated state predicted, is different. It is clear that, the smallest intervals  of $p_{pf}$ and $\theta$ in which the entangled behavior of the accelerated state is displayed when it passes through the phase channel $\mathcal{C}_{pf}$.

Finally, the effect of the  phase flip channel on the Wigner function is displayed  in Fig.(\ref{fig:3.3.5}), it has a strong effect on the coherence of the quantum correlation, where the upper hemisphere, which predicts the classical correlations, increases  as the channel strength $p_{pf}$ increases, where the classical correlations are displayed in $0\leq z\leq 1$, while the quantum correlations are described by a small lune.

\section{summary.}\label{s.3.6}

In this contribution, we investigate the Wigner function distribution of  accelerated and non-accelerated state, which is initially prepared in a maximum entangled state.  The robustness coherence  of this system  against different noisy channels is discussed, where  this decoherence is depicted  either  due to the noisy channels or to the acceleration process. The negative values of the Wigner function represent an indicator of the presence of the quantum correlation, while the positivity of Wigner function means that the system contains classical correlation.

The effect of the channels strengths, distribution angles, and the acceleration parameter on the behavior of the Wigner function  is investigated.  The general behavior of the Wigner function shows that, the separability (inseparability) of the accelerated  system  are shown in different intervals of these parameters. The minimum (maximum) values of the Wigner function  depend on the type of the noisy channel.  Since, we start with an entangled state, then the non-accelerated state has  negative values over all the distribution angles.

The behavior of the Wigner function on the surface of Bloch sphere is displayed at different values of channel strengths and the acceleration parameter. Different views of the   classical  and quantum correlations  are exhibited, cap, lower(upper) hemispheres and lower(upper) lune. The amplitude and the phase channels increase the classical correlation in $z-$ directions, while the bit and the phase bit channels depict the quantum correlation on $y$ and $z$ directions,respectively.

In general the Wigner function increases as the strength of any noisy channel increases. The distribution angles may be used as control parameters to suppress the decoherence of the initial quantum correlations.  For the amplitude damping channel, at particular values of the acceleration,  the robustness of quantum correlation is shown either at small values of the channel strength and  large values of the distribution angles or large values of the channel strength and small values of the distribution angles. The quantum correlation is displayed at small values of the channel strength and large values of the distribution angle, if the accelerated state passes through the bit flip channel. For the phase channel, these quantum correlations are independent of the channel strength if we set large values of the distribution angles.


\begin{thebibliography}{99}
\bibitem{PhysRevA.48.2479}
H\'ector Moya-Cessa and Peter~L. Knight.
\newblock Series representation of quantum-field quasiprobabilities.
\newblock {\em Phys. Rev. A}, 48:2479--2481, 1993.

\bibitem{deleglise2008reconstruction}
S.~Deleglise, I.~Dotsenko, C.~Sayrin, J.~Bernu, M.~Brune, J-Michel Raimond, and
  S.~Haroche.
\newblock Reconstruction of non-classical cavity field states with snapshots of
  their decoherence.
\newblock {\em Nature}, 455(7212):510, 2008.

\bibitem{mcconnell2015entanglement}
Robert McConnell, Hao Zhang, Jiazhong Hu, Senka {\'C}uk, and Vladan
  Vuleti{\'c}.
\newblock Entanglement with negative {W}igner function of almost 3,000 atoms
  heralded by one photon.
\newblock {\em Nature}, 519(7544):439, 2015.

\bibitem{mohamed2018nonclassical}
A.-B.A. Mohamed and N.~Metwally.
\newblock Nonclassical features of two {SC}-qubit system interacting with a
  coherent {SC}-cavity.
\newblock {\em Phys. E}, 102:1--7, 2018.

\bibitem{obada2012wigner}
A.-S.F. Obada, H.A. Hessian, A.-B.A. Mohamed, and M.~Hashem.
\newblock Wigner function and phase properties for a two-qubit field system
  under pure phase noise.
\newblock {\em J Russ. Laser Research}, 33(4):369--378, 2012.

\bibitem{biedenharn1984angular}
Lawrence~C. Biedenharn and James~D. Louck.
\newblock {\em Angular momentum in quantum physics: theory and application}.
\newblock Cambridge University Press, 1984.

\bibitem{klimov2017generalized}
A.~B. Klimov, J.~Romero, and H.~Guise.
\newblock Generalized {SU}(2) covariant {W}igner functions and some of their
  applications.
\newblock {\em J. Phys. A}, 50(32):323001, 2017.

\bibitem{husimi1940some}
K.~Husimi.
\newblock Some formal properties of the density matrix.
\newblock {\em Proceedings of the Physico-Mathematical Society}, 22:264--314,
  1940.

\bibitem{PhysRevA.57.671}
G.~S. Agarwal.
\newblock State reconstruction for a collection of two-level systems.
\newblock {\em Phys. Rev. A}, 57:671--673, 1998.

\bibitem{varilly1989moyal}
Joseph~C V{\'a}rilly and Jos{\'e}M Gracia-Bond{\'\i}a.
\newblock The {M}oyal representation for spin.
\newblock {\em Ann. phys.}, 190(1):107--148, 1989.

\bibitem{klimov20022}
A.B. Klimov and S.M. Chumakov.
\newblock On the {SU} (2) {W}igner function dynamics.
\newblock {\em Revista mexicana def{\'\i}sica}, 48(4):317--324, 2002.

\bibitem{PhysRevLett.10.277}
E.~C.~G. Sudarshan.
\newblock Equivalence of semiclassical and quantum mechanical descriptions of
  statistical light beams.
\newblock {\em Phys. Rev. Lett.}, 10:277--279, 1963.

\bibitem{PhysRevA.70.062101}
Kathleen~S. Gibbons, Matthew~J. Hoffman, and William~K. Wootters.
\newblock Discrete phase space based on finite fields.
\newblock {\em Phys. Rev. A}, 70:062101, 2004.

\bibitem{reboiro2015use}
M~Reboiro, O~Civitarese, and D~Tielas.
\newblock Use of discrete {W}igner functions in the study of decoherence of a
  system of superconducting flux-qubits.
\newblock {\em Phys. Scripta}, 90(7):074028, 2015.

\bibitem{ciampini2017wigner}
M.~A. Ciampini, T.~Tilma, M.~J. Everitt, W.J. Munro, P.~Mataloni, K.~Nemoto,
  and M.~Barbieri.
\newblock Wigner function reconstruction of experimental three-qubit {GHZ} and
  {W} states.
\newblock {\em arXiv preprint arXiv:1710.02460}, 2017.

\bibitem{PhysRevLett.117.180401}
T.~Tilma, M.~J. Everitt, J.~H. Samson, W.~J. Munro, and Kae Nemoto.
\newblock Wigner functions for arbitrary quantum systems.
\newblock {\em Phys. Rev. Lett.}, 117:180401, 2016.

\bibitem{koczor2018time}
B.~Koczor, R.~Zeier, and Steffen~J. Glaser.
\newblock Time evolution of coupled spin systems in a generalized {W}igner
  representation.
\newblock {\em Ann. Phys.}, Accepted Manuscript.

\bibitem{PhysRevA.24.2889}
G.~S. Agarwal.
\newblock Relation between atomic coherent-state representation, state
  multipoles, and generalized phase-space distributions.
\newblock {\em Phys. Rev. A}, 24:2889--2896, 1981.

\bibitem{metwally2019wigner}
N.~Metwally, M.Y. Rabbou, M.M.A. Ahmed, and A.-S.F. Obada.
\newblock Wigner function of accelerated and non-accelerated {G}reenberger
  {H}orne {Z}eilinger state.
\newblock {\em arXiv preprint arXiv:1901.08828}, 2019.

\bibitem{PhysRevA.82.042332}
David~E. Bruschi, Jorma Louko, Eduardo Mart\'{\i}n-Mart\'{\i}nez, Andrzej
  Dragan, and Ivette Fuentes.
\newblock Unruh effect in quantum information beyond the single-mode
  approximation.
\newblock {\em Phys. Rev. A}, 82:042332, 2010.

\bibitem{PhysRevA.83.052306}
Eduardo Mart\'{\i}n-Mart\'{\i}nez and Ivette Fuentes.
\newblock Redistribution of particle and antiparticle entanglement in
  noninertial frames.
\newblock {\em Phys. Rev. A}, 83:052306, 2011.

\bibitem{metwally2017estimation}
N.~Metwally.
\newblock Estimation of teleported and gained parameters in a non-inertial
  frame.
\newblock {\em Laser Physics Letters}, 14(4):045202, 2017.

\bibitem{PhysRevA.78.022322}
A.~Salles, F.~de~Melo, M.~P. Almeida, M.~Hor-Meyll, S.~P. Walborn, P.~H.
  Souto~Ribeiro, and L.~Davidovich.
\newblock Experimental investigation of the dynamics of entanglement: Sudden
  death, complementarity, and continuous monitoring of the environment.
\newblock {\em Phys. Rev. A}, 78:022322, 2008.

\bibitem{PhysRevA.87.042108}
B.~Horst, K.~Bartkiewicz, and A.~Miranowicz.
\newblock Two-qubit mixed states more entangled than pure states: Comparison of
  the relative entropy of entanglement for a given nonlocality.
\newblock {\em Phys. Rev. A}, 87:042108, 2013.

\bibitem{nielsen2002quantum}
M.~A. Nielsen and I.~Chuang.
\newblock {\em Quantum computation and quantum information}.
\newblock Cambridge University Press, Cambridge, 2002.

\end{thebibliography}

\end{document}